RESEARCH ARTICLE

# DYNAMICAL EVOLUTION OF THE OPEN CLUSTERS COLLINDER 110 AND NGC 2477

**Zahra AL[1,*], Yüksel KARATAŞ[2], Forough RAJAEI [3]**

[1] Institute of Graduate Studies in Sciences, İstanbul University, İstanbul, Turkey
zahraa96al@gmail.com - 0009-0004-4134-3752

[2] Department of Astronomy and Space Sciences, Faculty of Science, İstanbul University, İstanbul, Turkey
karatas@istanbul.edu.tr - 0000-0002-7737-6589

[3] Department of Physics, Faculty of Science, Yeditepe University, İstanbul, Turkey
forough.rajaei@std.yeditepe.edu.tr - 0009-0005-6440-0390

**Abstract**

Using Gaia DR3 photometric and astrometric data, we identified the likely members of the open clusters Collinder 110 and NGC 2477, and derived their astrophysical and dynamical parameters. We obtained reddening values of $E(B-V) = 0.43 \pm 0.05$ and $0.33 \pm 0.04$, and ages of $1.82 \pm 0.10$ Gyr and $1.05 \pm 0.10$ Gyr for Collinder 110 and NGC 2477, respectively. Within the uncertainties, the distances derived from the color–magnitude diagrams ($2.09 \pm 0.06$ and $1.38 \pm 0.04$ kpc) are consistent with those obtained from Gaia DR3 parallaxes ($2.08 \pm 0.31$ and $1.45 \pm 0.11$ kpc). Both clusters exhibit large core and limiting radii. The core radius ($R_c$) of Collinder 110 does not follow the expected decreasing trend with increasing evolutionary parameter ($\log \tau_2$). In addition to losing part of its low-mass stellar content, its relatively large $R_c$ may reflect primordial conditions and/or the presence of a possible stellar-mass black hole. The relatively large $R_c/R_h$ ratios, together with their advanced dynamical evolution ($\log \tau_2$ = 1.24 and 1.34), indicate that significant two-body relaxation and mass segregation have occurred within their central regions. The small values of $R_h/R_j$ (0.32 and 0.16) suggest that low-mass stars have dynamically redistributed within the comparatively large Jacobi radii under the combined influence of internal and external processes. However, their large $R_t/R_j$ ratios (1.60 and 1.23) imply that both clusters are tidally affected and have likely lost part of their low-mass stellar populations to the Galactic field. Despite these effects, their locations in the third Galactic quadrant, where giant molecular clouds are relatively scarce, may explain their survival to their present ages of 1.82 and 1.05 Gyr, respectively.

## 1. INTRODUCTION

Understanding the properties of open clusters (OCs) is crucial for investigating star formation and Galactic structure [1]. Stars in OCs form within the same molecular cloud and share a common origin, which makes them similar in fundamental properties such as age, distance, velocity, and proper motion. Gaia DR3 provides precise astrometric and photometric data for hundreds of open clusters, enabling reliable identification of cluster members within the surrounding field [2]. In particular, Gaia proper motion measurements allow accurate separation of cluster members from field stars, providing a robust basis for structural and dynamical analyses.

*Corresponding Author: zahraa96al@gmail.com

Investigating the dynamical properties of open clusters is essential for tracing their evolution under the combined influence of internal mechanisms — such as energy equipartition and core relaxation — and external effects, including tidal forces from the Galactic disk, spiral arms, the overall Galactic potential, and encounters with giant molecular clouds [3–8]. Diagnostic parameters such as the relaxation time, mass function slope, evolutionary parameter, core-to-half-mass radius ratio, and half-mass-to-Jacobi radius ratio help determine whether a cluster is expanding, approaching tidal filling, or progressing toward core collapse, thereby constraining its evolutionary stage and long-term stability. When a cluster becomes tidally filled, internal dynamical processes dominate its central evolution. Two-body relaxation, mass segregation, and core contraction reduce the core radius, while the half-mass radius generally remains nearly constant. Furthermore, primordial binaries and possible stellar-mass black holes in the core may inject energy into the system, contributing to cluster expansion. These structural changes are typically accompanied by enhanced mass loss from the outer regions. Under the influence of the Galactic tidal field, stars gain kinetic energy, which increases the evaporation rate. Over time, the combined action of these processes leads to the gradual dissolution of the cluster into the Galactic field.

Collinder 110 and NGC 2477 are located in the third Galactic quadrant at low Galactic latitudes, placing them close to the Galactic disk (see Figure 1 and Table 1). Both clusters have been previously studied using Gaia DR3 data [9, 10, 11, 8]. In this work, we employ the pyUPMASK algorithm on Gaia DR3 astrometric data to derive membership probabilities for cluster stars. The astrophysical parameters — including reddening, distance modulus (or distance), and age — are obtained through isochrone fitting to Gaia DR3 color–magnitude diagrams. For the isochrones, we adopt the spectroscopic metallicities ($[Fe/H]$) reported by [12] for Collinder 110 and [13] for NGC 2477. Gaia data are also used to determine the structural parameters, stellar masses, mass function slopes, dynamical quantities, and kinematic properties of the clusters. Based on these results, the dynamical evolution of the clusters is interpreted, and the relative importance of internal and external dynamical processes is assessed.

The structure of this paper is as follows. In Section 2, we describe the Gaia DR3 data, the determination of structural parameters (cluster dimensions), and the membership selection method. Section 3 presents the derivation of astrophysical parameters from isochrone fitting. Sections 4 and 5 focus on the kinematic properties, orbital parameters, cluster masses, and mass function analysis. Dynamical parameters and related indicators are discussed in Section 6. Finally, the discussion and conclusions, including comparisons with previous studies and interpretations of the dynamical evolution, are presented in the last section.

**Table 1.** The equatorial and Galactic coordinates of Collinder 110 and NGC 2477, taken from WEBDA.

| Parameter | Collinder 110 | NGC 2477 |
|---|---|---|
| $\alpha_{2000}$ $(h\ m\ s)$ | 06 38 24 | 07 52 10 |
| $\delta_{2000}$ $(^{\circ}\ '\ '')$ | +02 01 00 | -38 31 48 |
| $\ell$ (°) | 209.649 | 253.563 |
| $b$ (°) | -1.978 | –5.838 |

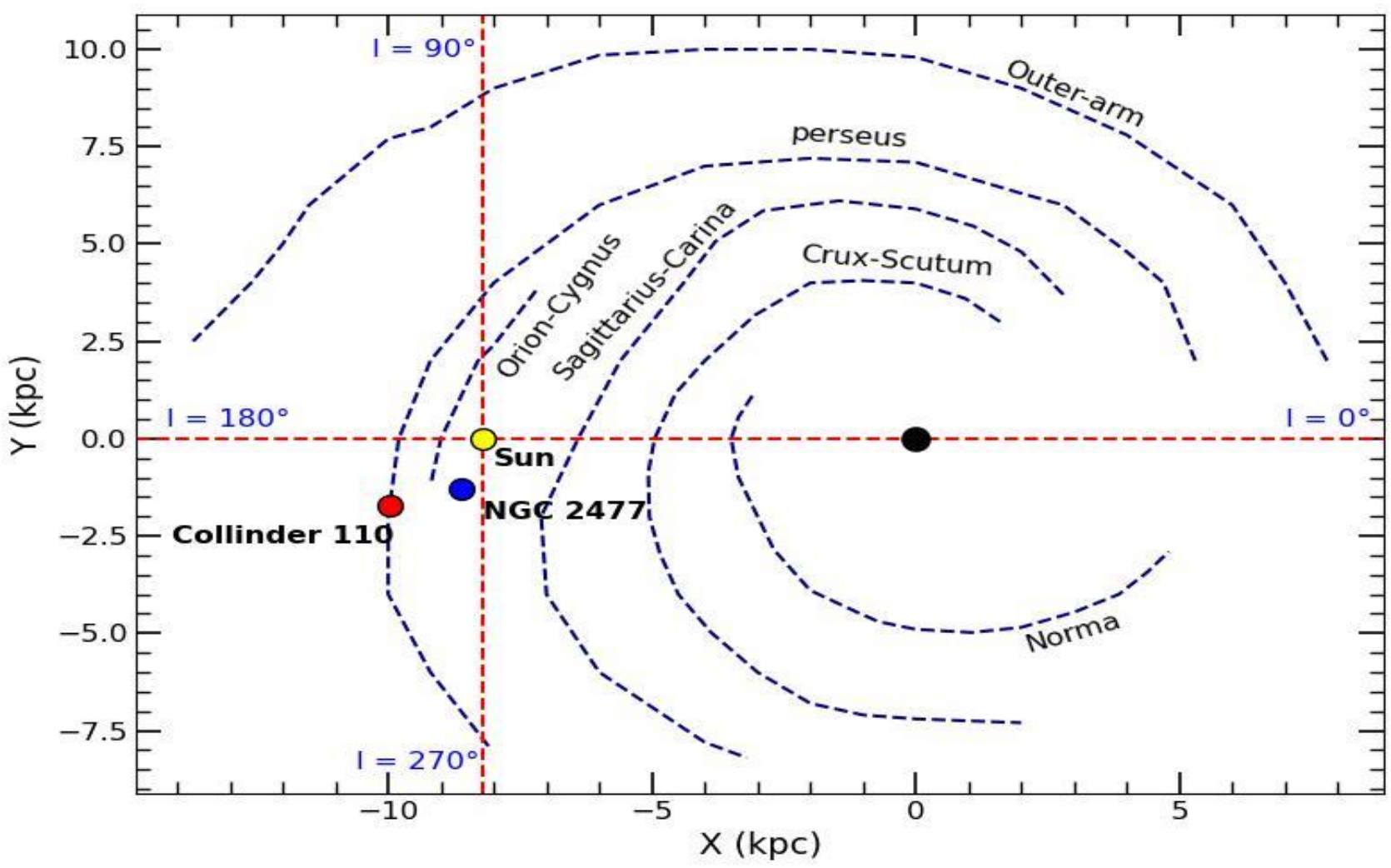

**Figure 1.** Spatial distribution of Collinder 110 (red circle) and NGC 2477 (blue circle) in the Galactic plane.. The Sun is indicated by a yellow circle at a Galactocentric distance of 8.2 ± 0.1 kpc (Bland-Hawthorn and Gerhard, 2016) from galactic center (black filled circle). The background image[1] is adapted from the original illustration by Robert Hurt (IPAC) and Bill Saxton (NRAO/AUI/NSF).

## 2. DATA, STRUCTURAL PARAMETERS AND MEMBERSHIP ANALYSIS

Gaia DR3 provides highly precise astrometric and photometric measurements for nearly 1.3 billion sources, including celestial coordinates (α, δ), trigonometric parallaxes ($\varpi$ (mas)), and proper motions in right ascension and declination ($\mu_\alpha$, $\mu_\delta$ (mas yr$^{-1}$)). The Gaia broad-band photometry ($G, G_{\rm BP}, G_{\rm RP}$) covers the magnitude range $3 \leq G \leq 21$ [2]. For Collinder 110 and NGC 2477, astrometric and photometric data, together with available radial velocities, were retrieved from the Gaia ESA Archive[2] using the *gaiadr3.gaia_source* table. Stellar atmospheric parameters — effective temperature ($T_{\rm eff}$), metallicity ($[M/H]$, $[Fe/H]$), and surface gravity ($\log g$) — were extracted from the *gaiadr3.astrophysical_parameters* table. These include estimates from the GSP-Spec module based on higher-resolution Radial

Velocity Spectrometer (RVS) spectra [14] and from the ESP-HS module [15], which provides $T_{\rm eff}$ and log g determinations. Using the cluster center coordinates listed in Table 1, all Gaia DR3 sources within a radius of 60′ were selected, yielding initial samples of 138,282 stars for Collinder 110 and 272,680 stars for NGC 2477. Following [16], quality constraints of RUWE < 1.4, $G_{\rm RP}$ < 18 mag, and $\varpi$ > 0 were applied, reducing the samples to 19,401 and 35,202 stars, respectively. Gaia DR3 parallaxes were corrected for the global zero-point offset following Lindegren et al., 2021[3]. In this analysis, $G_{\rm RP}$ magnitudes were preferred due to calibration issues affecting the G band. The resulting datasets were used to derive structural parameters and determine cluster membership.

To estimate cluster boundaries, radial density profiles (RDPs) were constructed (Figure 2). The King (1962) model [17], which effectively describes both inner and outer cluster regions, was fitted to the data (Eqs. 1–2). From these fits, the core radius ($R_{\rm c}$), limiting radius ($R_{\rm lim}$), tidal radius ($R_{\rm t}$), central stellar density ($\sigma_0$), and background density ($\sigma_{\rm bg}$) were derived. The results are listed in Table 2 along with comparisons to previous studies. The best-fit models yield reduced chi-square values of $\chi^2_{\rm red}$ = 0.69 for both clusters. The core radius ($R_{\rm c}$) corresponds to the distance at which the stellar density decreases to half of its central value. The limiting radius ($R_{\rm lim}$), often referred to as the King radius, represents the observational truncation boundary determined by the radial distribution of cluster members relative to the field. The tidal radius ($R_{\rm t}$) defines the distance at which the cluster's gravitational potential balances the external Galactic tidal field, marking the point where the stellar density approaches the background level.

[1] https://www.universetoday.com/articles/our-place-in-the-galactic-neighborhood-just-got-an-upgrade

[2] https://gea.esac.esa.int/archive

[3] https://gitlab.com/icc-ub/public/gaiadr3-zeropoint

$$\sigma_{\rm R} = \begin{cases} \sigma_{\rm bg} + k\left[\frac{1}{\sqrt{1+(\frac{R}{R_{\rm c}})^2}} - \frac{1}{\sqrt{1+(\frac{R_{\rm t}}{R_{\rm c}})^2}}\right]^2 & , R \leq R_{\rm t} \\ \sigma_{\rm bg} & , R > R_{\rm t} \end{cases} \tag{1}$$

$$k = \sigma_0\left[1 - \frac{1}{\sqrt{1+(\frac{R_{\rm t}}{R_{\rm c}})^2}}\right]^{-2} \tag{2}$$

**Table 2.** Structural parameters of Collinder 110 and NGC 2477. References are given in the final column. N/A indicates values not available in the cited literature.

| Cluster | (1 pc) | $R_{\rm c}$ | $R_{\rm lim}$ | $R_{\rm t}$ | $R_{\rm c}$ | $R_{\rm lim}$ | $R_{\rm t}$ | Reference |
|---|---|---|---|---|---|---|---|---|
| | (1′) | (′) | (′) | (′) | (pc) | (pc) | (pc) | |
| (1) | (2) | (5) | (6) | (7) | (10) | (11) | (12) | (13) |
| Collinder 110 | 1.67 | 7.73±0.72 | 41.2±3.55 | 45.00±13.45 | 4.63±0.43 | 24.67±2.13 | 26.95±8.05 | This paper |
| | N/A | 9.61 | N/A | 22.37 | 6.43 | N/A | 14.95 | [18] |
| | N/A | $6.73^{0.24}_{0.27}$ | $28.56^{2.65}_{3.02}$ | N/A | $4.50^{0.16}_{0.18}$ | $19.10^{1.77}_{2.02}$ | N/A | [11] |
| | N/A | N/A | 27.50 | N/A | N/A | 18.33 | N/A | [19] |
| | N/A | 9.07±1.37 | N/A | 43.22±9.07 | 5.43±0.82 | N/A | 25.88±5.43 | [7] |
| | N/A | N/A | N/A | 30.98±2.99 | N/A | N/A | 20.65±1.99 | [20] |
| | N/A | N/A | 22.00±2.00 | N/A | N/A | 14.67±1.33 | N/A | [21] |
| NGC 2477 | 2.38 | 5.59±0.35 | 52.70±2.56 | 55.00±21.54 | 2.35±0.15 | 22.19±1.08 | 23.16±9.07 | This paper |
| | N/A | 14.99 | 72.79 | 20.79 | 6.04 | 29.31 | 8.37 | [18] |
| | N/A | $4.37^{0.26}_{0.19}$ | $109.04^{77.09}_{49.79}$ | N/A | $1.84^{0.11}_{0.08}$ | $45.97^{32.50}_{20.99}$ | N/A | [11] |
| | N/A | N/A | 33.68 | N/A | N/A | 14.21 | N/A | [19] |
| | N/A | N/A | N/A | N/A | N/A | N/A | N/A | [7] |
| | N/A | N/A | N/A | 36.67±2.73 | N/A | N/A | 15.47±1.15 | [20] |

Central density $\sigma_{0\rm K}$ and background density $\sigma_{\rm bg}$ are determined as 4.82±0.66, 1.63±0.03 star/arcmin and 14.68±1.30, 1.63±0.03 star/arcmin for Collinder 110 and NGC 2477, respectively. Within the determined limiting radii, $R_{\rm lim} = 27.56\ pc$ for Collinder 110 and $R_{\rm lim} = 22.22\ pc$ for NGC 2477, we determined membership probability (P) for each star.

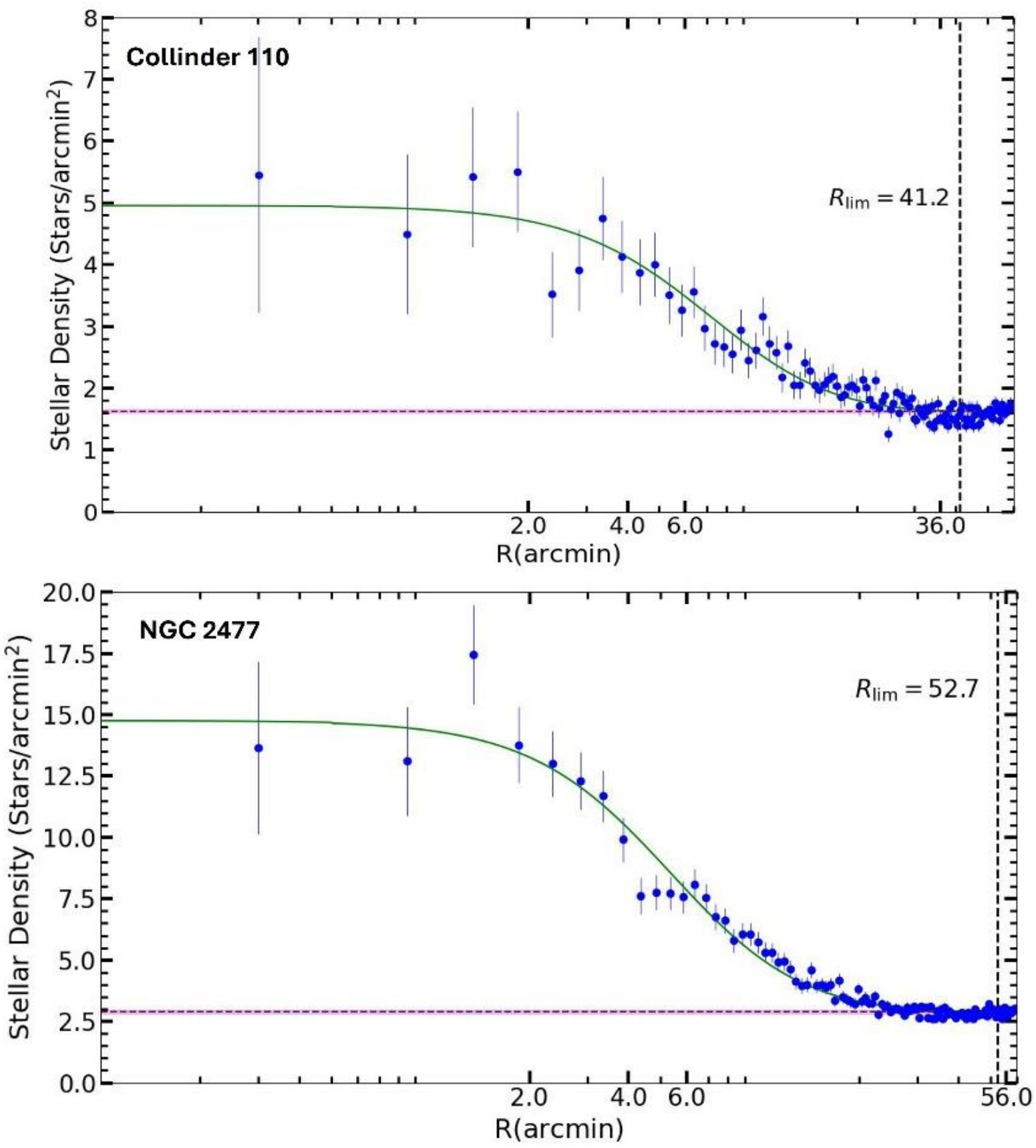

**Figure 2.** The radial density profiles (RDP) for 19401 stars in the region of Collinder 110 (top panel) and 35202 stars in the region of NGC 2477 (bottom panel). Solid lines and horizontal lines represent the best King 1962 fits and stellar background levels measured in the comparison field, respectively. Errors of the dots are from $\sqrt{N}$, where N is the number of stars used in the density estimation.

To determine cluster membership within $R_{\mathrm{lim}}$, we applied the pyUPMASK[4] algorithm with k-means clustering method [22] to Gaia DR3 astrometric data. pyUPMASK is a Python implementation of UPMASK (Unsupervised Photometric Membership Assignment in Stellar Clusters) [23]. From the histograms of membership probabilities for both clusters (Figure 3), we adopted the first significant rise in the probability distribution (P > 80%) as the membership threshold. This criterion yielded 810 likely members for Collinder 110 and 2050 for NGC 2477. The most distant member selected by pyUPMASK lies 39.28 arcmin from the center of Collinder 110, whereas the most distant member of NGC 2477 is located 16.37 arcmin from its center. Their proper motion diagrams, known as Vector Point Diagrams (VPDs), are presented in Figure 4. Squares centered on the mean proper motion components (first row of Table 3) are shown for both clusters. As seen in Figure 4, the probable cluster members exhibit a relatively tight concentration around the mean proper motion values compared to the more dispersed field stars.

The top right panel of Figure 4 represents proper motion velocity vectors plotted on the stars' equatorial positions $\alpha_{2000}$ versus $\delta_{2000}$. Each vector indicates the direction and magnitude of a star's movement across the sky and is color-coded according to the star's membership probability. Their CMDs (Figure 4) present an extended main sequence with some red giants, as expected for old OCs. In Figure 4, the evolved stars of NGC 2477 form a relatively dense group around the giant sequence, whereas two of the five evolved probable members of Collinder 110 are far away from the single giant sequence.

[4] https://github.com/msolpera/pyUPMASK

Their median Gaia DR3 astrometric parameters (proper motion components and parallaxes) for the likely members are listed in Table 3 together with the literature values. Within the uncertainties, the median values are compatible with those reported by [18, 24, 25].

Their Bayesian distances are 2079 ± 99 pc and 1447 ± 18 pc, respectively. The median equatorial coordinates ($\alpha_{2000}$, $\delta_{2000}$) of the probable members are 99.68°, 2.08° (Collinder 110) and 118.05°,-38.54° (NGC 2477), respectively, which are in good agreement with the coordinates in Table 1. The distribution of the equatorial coordinates of the likely members around each cluster's center is displayed in Figure 5.

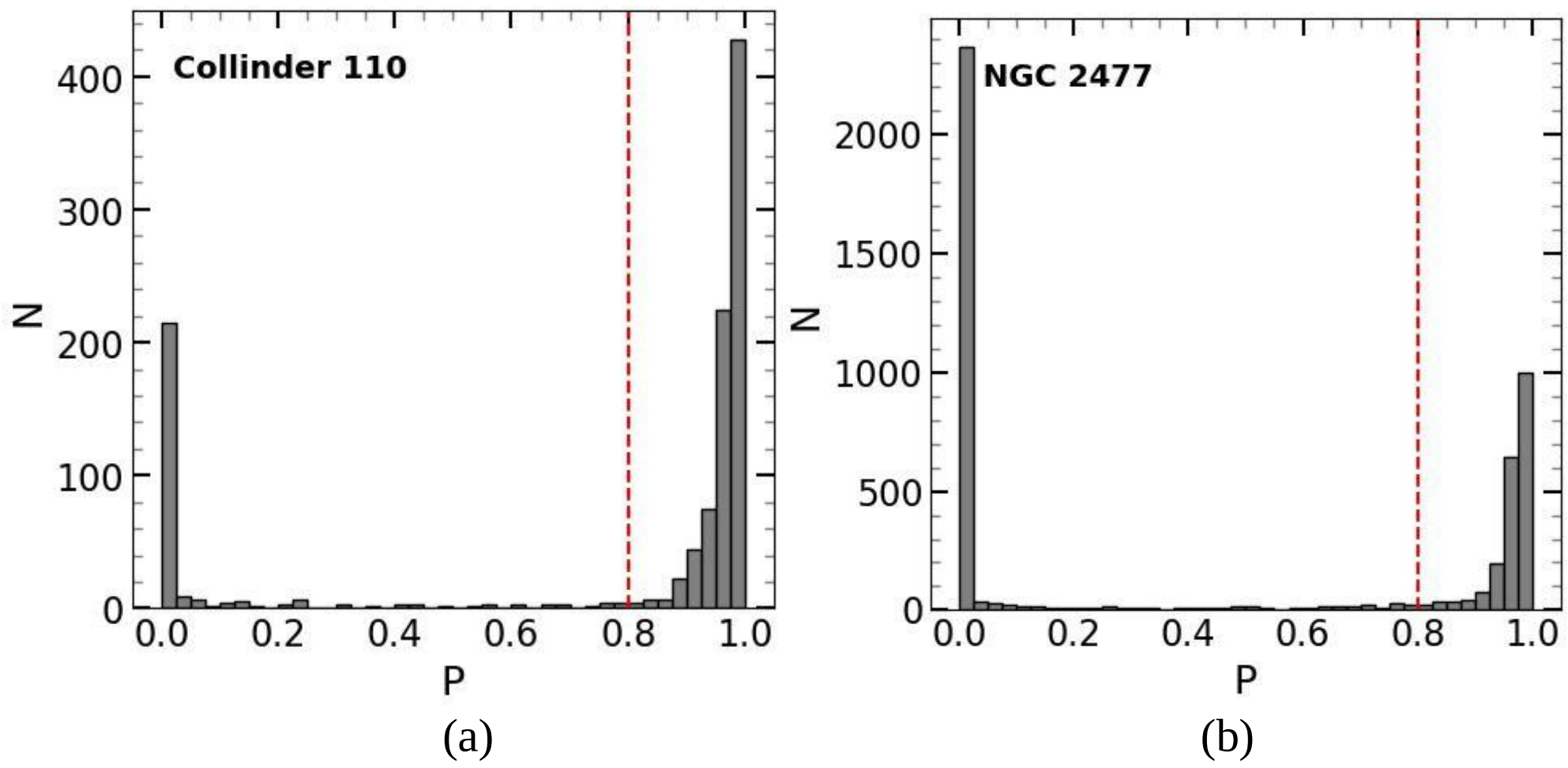

**Figure 3.** The histogram of membership probabilities according to pyUPMASK of Collinder 110 (panel (a)) and NGC 2477 (panel (b)). The vertical dotted lines show the selected probability limit value, 0.8.

**Table 3.** The median proper motion components, parallaxes and distances of the likely members. The values of [18, 24, 25] are listed in the next rows.

| Collinder 110 | | | | | NGC 2477 | | | | |
|---|---|---|---|---|---|---|---|---|---|
| $\mu_\alpha$ (mas/yr) | $\mu_\delta$ (mas/yr) | $\varpi$ (mas) | d (pc) | N | $\mu_\alpha$ (mas/yr) | $\mu_\delta$ (mas/yr) | $\varpi$ (mas) | d (pc) | N |
| -1.092±0.005 | -2.041±0.005 | 0.482±0.072 | 2075±310 | 810 | -2.428±0.004 | 0.906±0.005 | 0.691±0.054 | 1447±113 | 2050 |
| -1.093±0.003 | -2.039±0.003 | 0.435±0.002 | 2101 | 867 | -2.427±0.003 | 0.908±0.004 | 0.692±0.001 | 1384 | 2081 |
| -1.103±0.158 | -2.049±0.157 | 0.423±0.069 | 1991±80 | 925 | -2.455±0.186 | 0.866±0.211 | 0.667±0.040 | 1351±47 | 1870 |
| -1.091±0.143 | -2.049±0.144 | 0.425±0.063 | 2183 | 881 | -2.449±0.169 | 0.870±0.192 | 0.665±0.037 | 1442 | 1713 |

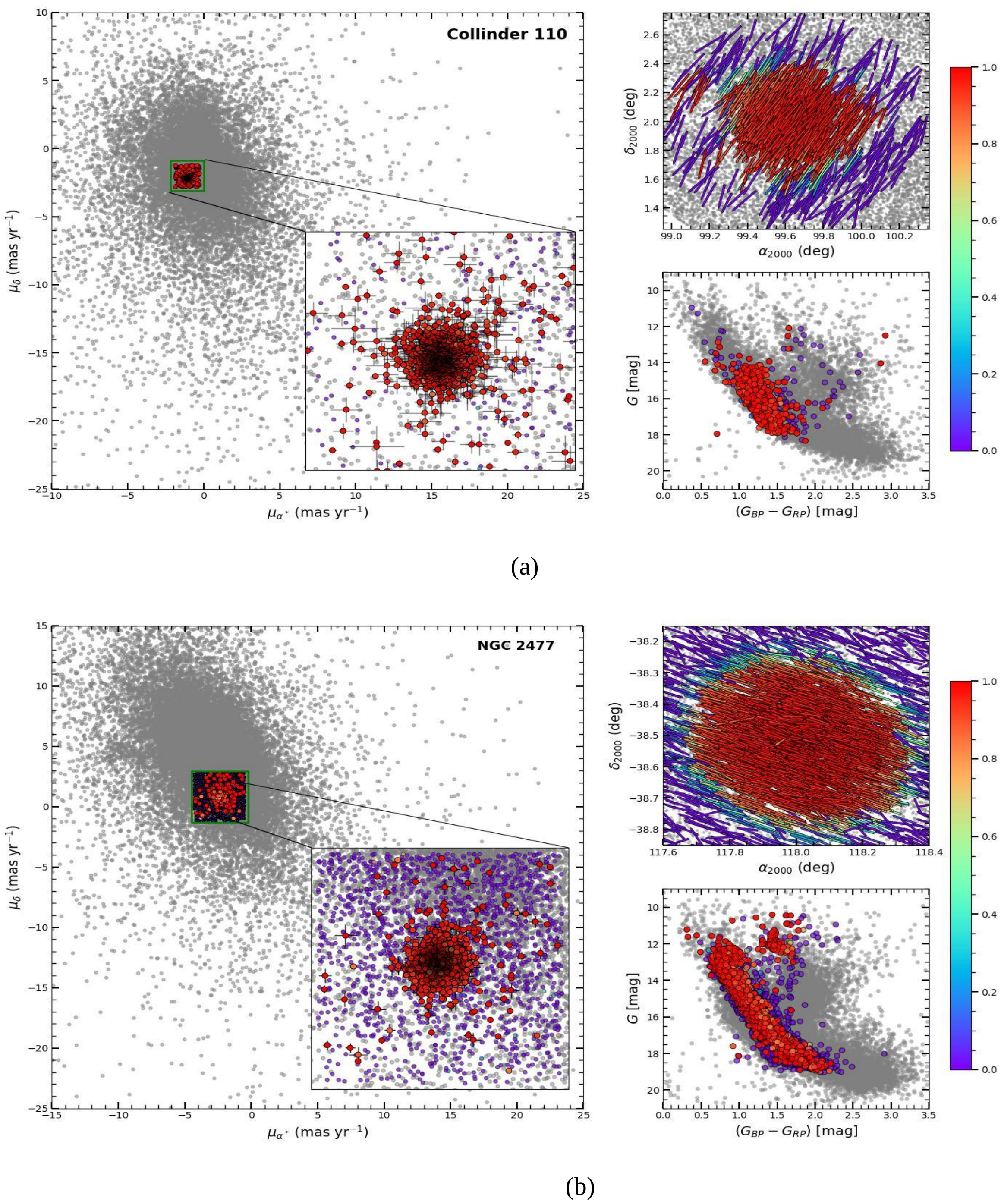

**Figure 4.** Astrometric and photometric analysis of Collinder 110 (panel a) and NGC 2477 (panel b). Each set displays the Vector-Point Diagram (VPD; left), proper-motion velocity vectors (top-right), and the $G$ vs. $(G_{BP}-G_{RP})$ color-magnitude diagram (CMD; bottom-right). Candidate members are color-coded by membership probability (see scale bar at far right), with field stars represented in gray. In the VPD, black bars in the inset denote median uncertainties in $\mu_\alpha$ and $\mu_\delta$. In the CMD, black error bars indicate representative uncertainties in $G$ and $G_{BP}-G_{RP}$. High-probability members (810 for Collinder 110; 2050 for NGC 2477) are plotted with larger symbols in the velocity-vector diagrams for enhanced visibility. All proper motions are presented in the Gaia reference frame.

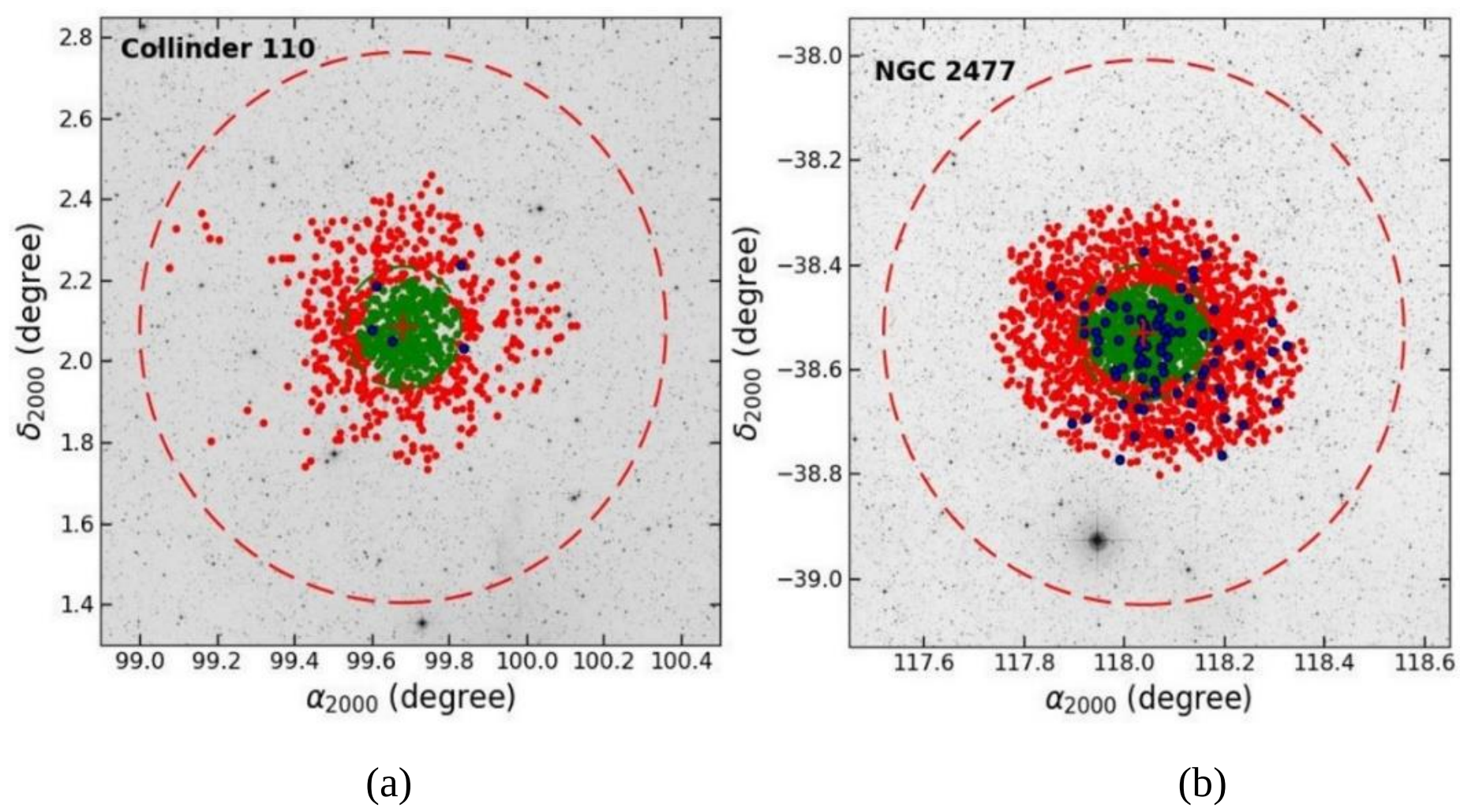

(a) (b)

**Figure 5.** The star charts of Collinder 110 (panel (a)) and NGC 2477 (panel (b)) from the webpage of The STScI Digitized Sky Survey. Red plus, green/red dots and green/red dashed circles represent the central equatorial coordinates, core/cluster members and core/cluster radius, respectively. Blue dots denote candidate evolved stars in the CMDs (Figure 6).

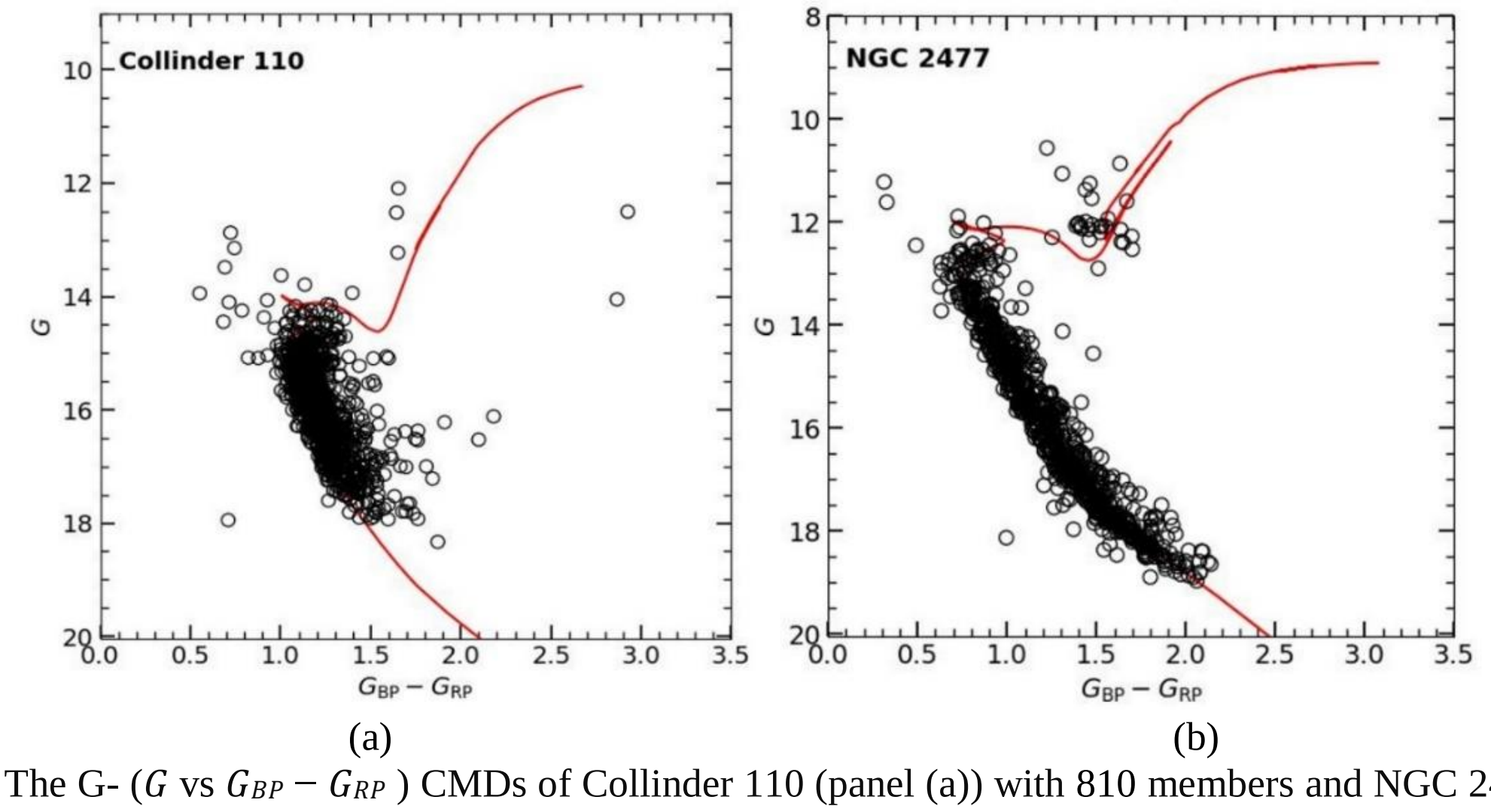

(a) (b)

**Figure 6.** The G- ($G$ vs $G_{BP} - G_{RP}$ ) CMDs of Collinder 110 (panel (a)) with 810 members and NGC 2477 (panel (b)) with 2050 members.

## 3. ASTROPHYSICAL PARAMETERS

To determine the distance moduli and ages of Collinder 110 and NGC 2477, PARSEC isochrones [26] were fitted to the member stars in the $G - G_{BP-RP}$ CMDs (Fig. 6). The two evolved candidate members of Collinder 110 were not used to constrain the isochrone fit because the red giant branch is represented by only two stars. The fit was therefore based primarily on the well-defined main-sequence population, while the evolved stars were retained as probable astrometric members. For the isochrone fitting, spectroscopic metallicities of $[Fe/H] = -0.10 \pm 0.04$ [12] for Collinder 110 and $[Fe/H] = +0.07 \pm 0.11$ [13] for NGC 2477 were adopted. These metallicities were converted to Z values via the relation $Z = Z_{\odot}10^{[\frac{\mathrm{Fe}}{\mathrm{H}}]}$, yielding $Z = 0.0121$ for Collinder 110 and $Z = 0.0179$ for NGC 2477. The PARSEC isochrones were been shifted both vertically and horizontally by $E(G_{BP-RP})$ on the CMDs (Figure 6) until the best fit to the observed intermediate section of the main sequence as well as the Red Giant/Red Clump sequences was obtained. This vertical shift is the true distance modulus, $DM_{\mathrm{o}} = (G_{\mathrm{o}}$ -$M_{\mathrm{G}})$. To determine the age, the PARSEC isochrones have also been shifted both vertically and horizontally in the CMDs to the amounts of $M_G + 2.85E(G_{BP-RP}) + DM$ and $(G_{BP-RP})_o$+ $E(G_{BP-RP})$, respectively. The derived astrophysical parameters (reddening, distance modulus/distance and age) are listed in Table 4. The $E(B-V)$ values were converted from $E(G_{BP-RP})$=0.55±0.06 (Collinder 110) and $E(G_{BP-RP})$ = 0.43±0.05 (NGC 2477) via $E(B-V) = 0.775E(G_{BP-RP})$. The derived ages are $1.82 \pm 0.10$ Gyr for Collinder 110 and $1.05 \pm 0.10$ Gyr for NGC 2477, respectively. The fitted isochrones (Figure 6.) gave distances as 2.09 kpc for Collinder 110 and 1.38 kpc for NGC 2477, respectively. The two evolved candidates of Collinder 110 that lie away from the giant sequence were not included in the isochrone fitting.

**Table 4.** Astrophysical parameters of Collinder 110 and NGC 2477 in this study and a comparison with the literature.

| Cluster | $E(B-V)$ | $(V_o - M_v)$ | $d$ (kpc) | $\log Age$ (age/yr) | Age (Gyr) | $[Fe/H]$ (dex) | Data source | Reference |
|---|---|---|---|---|---|---|---|---|
| Collinder 110 | 0.43±0.05 | 11.6±0.5 | 2.09±0.06 | 9.26±0.02 | 1.82±0.10 | -0.10±0.04 | Gaia DR3 | This study |
| | 0.53±0.02 | 11.50±0.09 | 1.99±0.08 | 9.09±0.24 | 1.23±0.91 | +0.03±0.02 | Gaia DR3 - 2MASS | [19] |
| | 0.52±0.10 | 11.47±0.30 | 1.97±0.27 | 9.25±0.10 | 1.78±0.46 | -0.13±0.23 | Gaia DR3 | [7] |
| | -- | -- | 1.95 | 9.23 | 1.70 | +0.03±0.02 | Gaia DR2 - Apogee | [27] |
| | 0.50 | -- | 2.20 | 9.15 | 1.41 | -- | Gaia DR2 | [28] |
| | -- | -- | 2.18 | 9.26 | 1.82 | -0.10±0.05 | Gaia ESO IDR6 | [29] |
| | -- | 11.70 | 2.18 | 9.26 | 1.82 | -- | Gaia EDR3 | [30] |
| | 0.53±0.02 | -- | 1.99±0.08 | 9.16±0.06 | 1.45±0.21 | +0.02±0.02 | Gaia DR2 | [24] |
| | 0.42 | -- | 2.36 | 9.22 | 1.66 | -- | PPMXL - 2MASS | [31] |
| | -- | -- | 1.95 | 9.15 | 1.41 | +0.01 | Spectrosccopy | [32] |
| | 0.50 | 11.45 | -- | -- | -- | -- | 2MASS | [33] |
| | 0.40 | 13.04 | 4.79 | 9.11 | 1.29 | +0.03±0.02 | Spectrosccopy | [34] |
| | 0.50 | -- | -- | 9.15 | 1.41 | -- | CCD UBVRI | [35] |
| NGC 2477 | 0.33±0.04 | 10.7±0.4 | 1.38±0.04 | 9.02±0.04 | 1.05±0.10 | +0.07±0.11 | Gaia DR3 | This study |
| | 0.38±0.03 | 10.65±0.07 | 1.35±0.05 | 8.93±0.10 | 0.85±0.22 | +0.07±0.03 | Gaia DR3 - 2MASS | [19] |
| | 0.40±0.10 | 10.59±0.30 | 1.31±0.18 | 9.05±0.10 | 1.22±0.19 | -0.13±0.23 | Gaia DR3 | [7] |
| | -- | -- | 1.44 | 9.05 | 1.22 | +0.07±0.04 | Gaia DR2 - Apogee | [27] |
| | 0.31 | -- | 1.44 | 8.85 | 0.71 | -- | Gaia DR2 | [28] |
| | -- | -- | 1.44 | 9.05 | 1.22 | +0.14±0.04 | Gaia ESO IDR6 | [29] |
| | -- | 10.80 | 1.44 | 9.05 | 1.22 | -- | Gaia EDR3 | [30] |
| | 0.38±0.03 | -- | 1.35±0.05 | 8.97±0.04 | 0.93±0.09 | +0.08±0.03 | Gaia DR2 | [24] |
| | 0.29 | -- | 1.45 | 8.92 | 0.83 | -- | PPMXL - 2MASS | [31] |
| | -- | -- | 1.34 | 8.85 | 0.71 | +0.11 | Spectrosccopy | [32] |
| | 0.23 | 11.55 | -- | -- | -- | -- | 2MASS | [33] |
| | 0.28 | -- | -- | 8.85 | 0.71 | -- | CCD UBVRI | [35] |

## 4. KINEMATICS AND ORBITAL PARAMETERS

The three-dimensional heliocentric velocity components $(U, V, W)$ for Collinder 110 and NGC 2477 were computed following the prescriptions of [36]. The calculations incorporated Gaia DR3 radial velocities (Table 5), together with the clusters' median proper motions and adopted distances. We adopted the Gaia DR3 photometric distances listed in Table 5. To place the velocities in the Local Standard of Rest (LSR) frame, the solar motion $(U, V, W)_\odot = (+11.10, +12.24, +7.25)$ km/s relative to the LSR [37] was subtracted, yielding the corrected components $(U', V', W')$. A Galactocentric distance of the Sun $R_\odot = 8.2 \pm 0.1$ kpc [38] and an LSR circular velocity of $V_{\rm LSR} = 239$ km/s [39] were adopted throughout the analysis. The heliocentric Cartesian positions $(x', y', z')$ and the LSR-corrected velocity components were subsequently transformed into the Galactic Standard of Rest (GSR) frame, providing the Galactocentric coordinates $(x, y, z)$ and velocity components $(V_x, V_y, V_z)$.

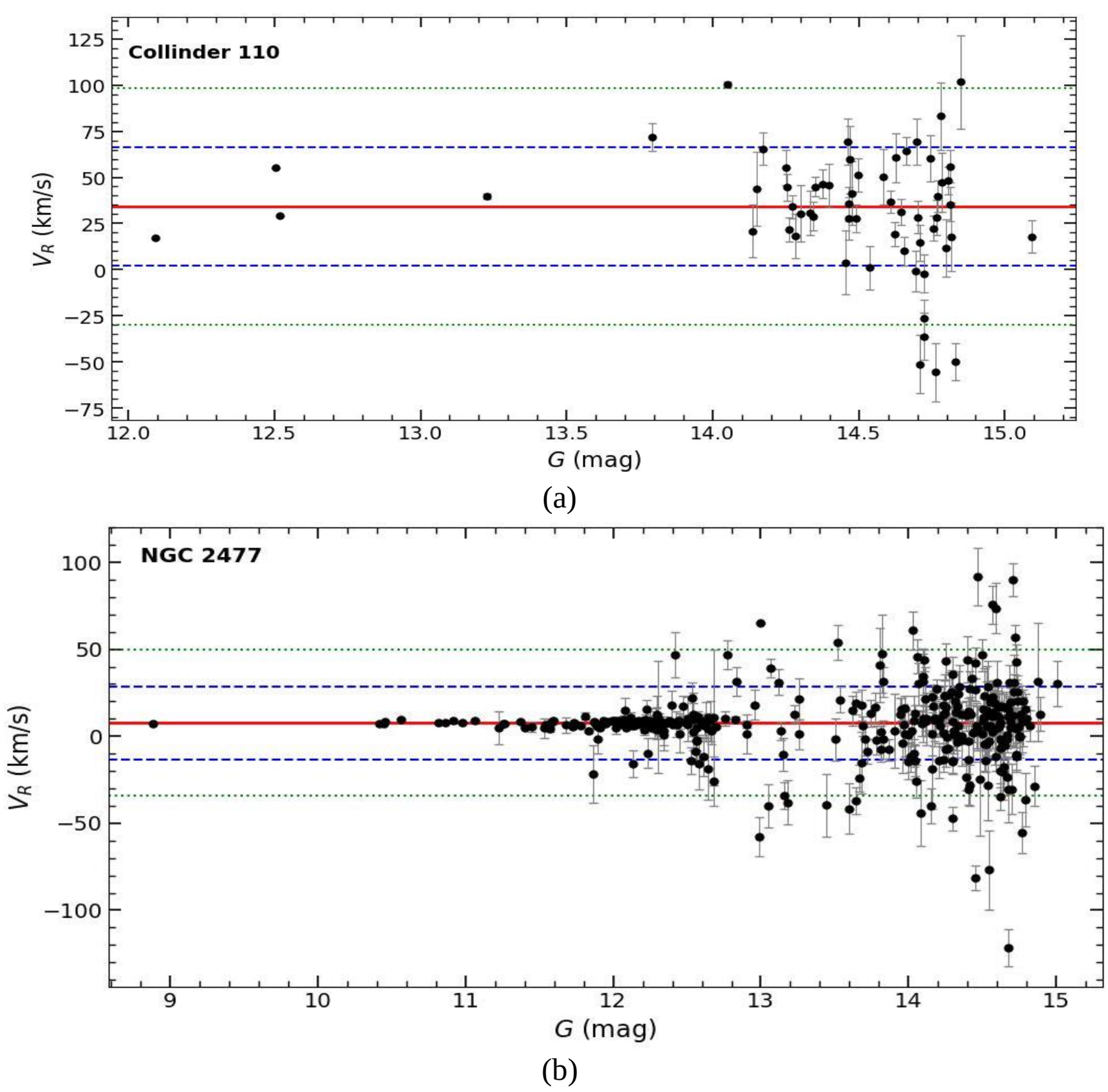

**Figure 7.** Radial velocity versus Gaia $G$- magnitude for the 56 members of Collinder 110 (panel (a)) and 324 members of NGC 2477 (panel (b)). The solid red line denotes the median radial velocity; $V_R = 34.15 \pm 5.24\ kms^{-1}$ for Collinder 110, and $V_R = 7.71 \pm 1.38\ kms^{-1}$ for NGC 2477. The blue dashed and the green dotted lines denote the ±1$\sigma$ and ±2$\sigma$ limits, respectively.

**Table 5.** The average radial velocity ($V_R$) km/s, space velocity components and rotational velocity ($U, V, W, V_\phi$) km/s, eccentricity (ecc) and peri- and apogalactic distances, initial and present-day distances ($R_{max}, R_{min}, R_m, z_{max}, R_{in}, R_{GC}(kpc)$. The orbital angular momentum ($J_z$) (kpc km/s). T is the time of one revolution around the Galactic center as Myr. $N_{Rev}$ is the number of revolutions over the age of the cluster.

| **Parameter** | **Collinder 110** | **NGC 2477** |
|---|---|---|
| $V_R$ $(km/s)$ | 34.15 ± 5.24 | 7.01 ± 1.67 |
| $U$ $(km/s)$ | -23.26 ± 4.53 | -14.39 ± 053 |
| $V$ $(km/s)$ | -28.20 ± 2.94 | -2.39 ± 1.31 |
| $W$ $(km/s)$ | -19.91 ± 1.90 | -11.43 ± 0.35 |
| $U_{LSR}(km/s)$ | -12.16 ± 4.53 | -3.29 ± 0.53 |
| $V_{LSR}(km/s)$ | -15.96 ± 2.94 | 9.85 ± 1.31 |
| $W_{LSR}(km/s)$ | -12.66 ± 1.90 | -4.18 ± 0.35 |
| $V_\varphi$ $(km/s)$ | 223.12 ± 2.66 | 246.48 ± 1.23 |
| *ecc* | 0.06 ± 0.01 | 0.12 ± 0.01 |
| $R_{min}$ *(kpc)* | 9.00 ± 0.18 | 8.10 ± 0.02 |
| $R_{max}$ *(kpc)* | 10.21 ± 0.12 | 10.25 ± 0.11 |
| $R_m$ *(kpc)* | 9.68 ± 0.01 | 9.09 ± 0.01 |
| $Z_{max}$ *(kpc)* | Q | 0.15 ± 0.01 |
| $R_{in}$ *(kpc)* | 7.69 ± 0.36 | 9.29 ± 0.08 |
| $R_{GC}$ *(kpc)* | 10.07 ± 0.04 | 8.69 ± 0.11 |
| $J_z$ $(kpcx\ km/s)$ | 2247.52 ± 27.75 | 2141.66 ± 11.70 |
| *T (Myr)* | 191.19 ± 0.01 | 178.18 ± 0.01 |
| $N_{rev}$ | 9.52 ± 0.49 | 5.17 ± 0.01 |

From the Gaia DR3 velocity versus G-mag (Figure 7), the median radial velocitiy is estimated as $V_R = 34.15 \pm 5.24\ \mathrm{kms}^{-1}$ for 56 members of Collinder 110, and $V_R = 7.71 \pm 1.38\ \mathrm{kms}^{-1}$ for 324 members of NGC 2477, respectively.

We calculated the azimuthal velocity component, $V_\phi$ $(km/s)$, using the following expression:

$$V_\phi = \frac{xV_y - yV_x}{R} \quad (3)$$

In this coordinate convention, a negative $V_\phi$ value signifies prograde motion. To determine the orbital characteristics—specifically the peri- and apo-galactic distances ($R_{min}$, $R_{max}$) and the maximum vertical displacement ($z_{max}$), we utilized the "MWPotential2014" model provided by the galpy library [40]. From these spatial parameters, the orbital eccentricity (ecc) was derived as follows:

$$ecc = \frac{R_{max} - R_{min}}{R_{max} + R_{min}} \quad (4)$$

The guiding radius (or mean galactocentric distance), denoted as $R_m$, was computed as the arithmetic mean of $R_{min}$ and $R_{max}$. Using the kinematic data and estimated ages presented in Tables 5 and 4, respectively, we integrated the clusters' orbits within the cumulative Galactic potential described by [40]. Furthermore, the vertical component of the angular momentum, $J_z$ (kpc km/s), was determined using $J_z = xV_y - yV_x$. The resulting orbital trajectories for Collinder 110 and NGC 2477 are visualized in Figure 8. These include the projection onto the Galactic (x-y) plane and the motion within the meridional (z-R) plane. All calculated orbital parameters are summarized in Table 5.

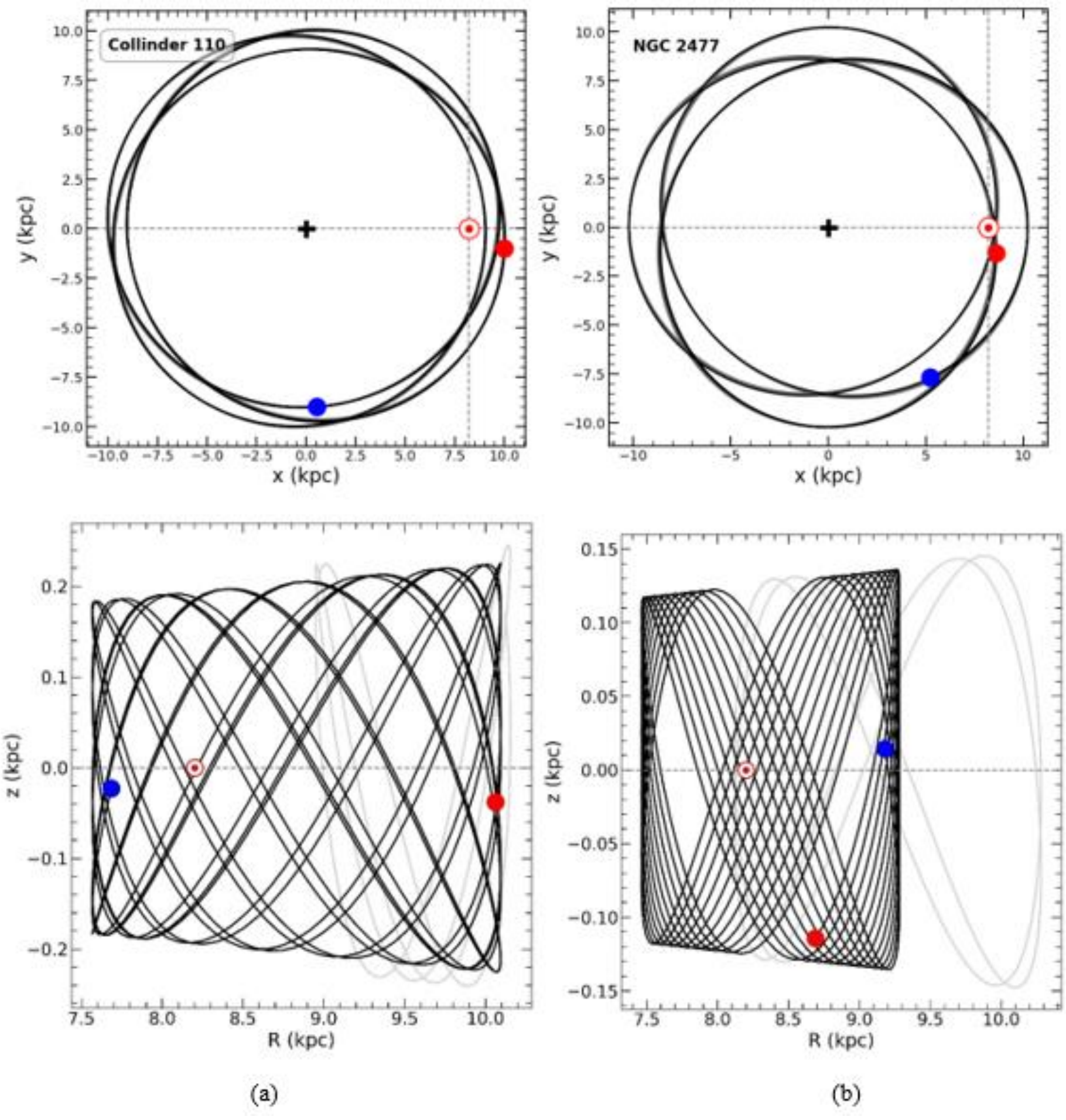

**Figure 8.** Galactic orbits of Collinder 110 (column (a)) and NGC 2477 (column (b)). The black trajectories represent the paths traveled by both OCs through their ages, while the gray trajectories represent the future paths of the OCs through a one Gyr. The filled blue/red dots show their initial/present day positions. The open red circle shows the Sun's position

## 5. MASS AND MASS FUNCTION SLOPE

From the PARSEC isochrones, we assigned individual masses to the 810 identified members of Collinder 110 and the 2050 members of NGC 2477. We converted the apparent $G$ magnitudes into absolute $M_{\rm G}$ magnitudes, accounting for the distance moduli and reddening values provided in Table 4. Based on these calculations, the aggregate masses for Collinder 110 and NGC 2477 were determined to be $1097.30 \pm 48.18\ M_{\odot}$ and $2512.95 \pm 10.42\ M_{\odot}$, respectively. These mass uncertainties were estimated by incorporating both the mass function errors and membership probabilities. The Mass Function (MF), which represents the stellar mass distribution within the cluster volume, was modeled using the following power-law relation:

$$\frac{dN}{dM} = \phi(M) = M^{-(1+\chi)} \tag{5}$$

Here $dN$ denotes the number of stars within a mass range $dM$ centered at mass $M$, while $\chi$ represents the slope of the mass function. To ensure data reliability, we restricted our sample to stars brighter than $G$ =19 mag, consistent with the Gaia DR3 completeness threshold [41]. This selection resulted in a refined sample of 799 members for Collinder 110 and 1220 members for NGC 2477 for the MF analysis. The resulting MF relations are given by $\log\frac{N}{\Delta M} = (-2.23 \pm 2.74) \times \log\frac{M}{M_{\odot}} + 0.83$ for Collinder 110, and $\log\frac{N}{\Delta M} = (-2.84 \pm 1.03) \times \log\frac{M}{M_{\odot}} + 0.83$ for NGC 2477, corresponding to MF slopes of $\chi$=1.23 for Collinder 110 and $\chi$ = 1.84 for NGC 2477, respectively. The steep MF for Collinder 110 is fully consistent with the canonical Kroupa IMF value of $\chi$ = 1.3 [42] within the uncertainties. Although the MF of NGC 2477 is slightly steep, it still lies within the range typically found for dynamically evolved open clusters, and therefore shows reasonable agreement with the Kroupa IMF.

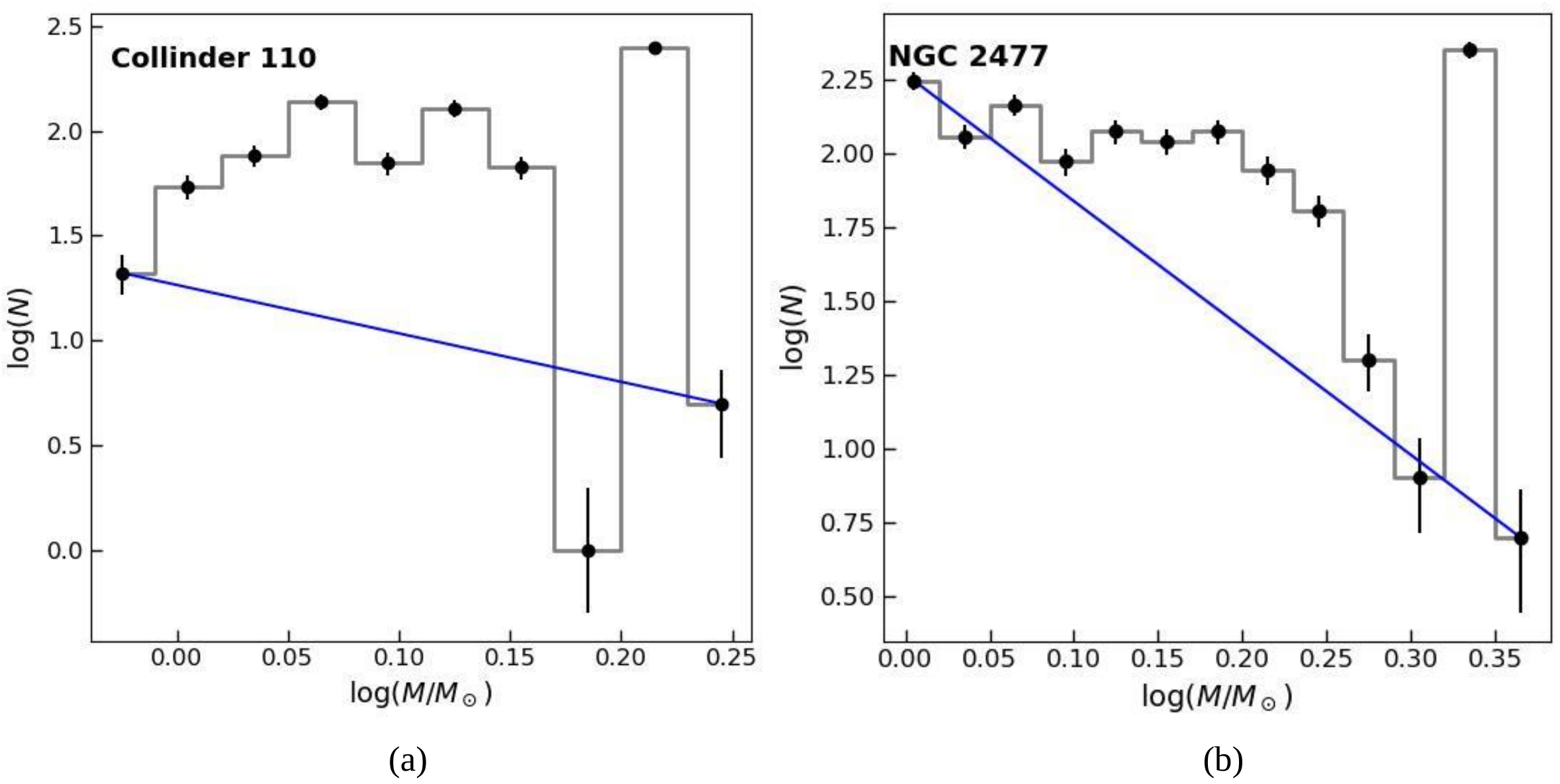

**Figure 9.** The mass function of Collinder 110 (panel (a)) and NGC 2477 (panel (b)). The vertical bars represent the poisson errors an are proportional to $\sqrt{N}$.

The half-mass radius ($R_{\rm h}$) represents the distance from the cluster's core that encompasses half of its total stellar mass. For our target systems, we calculated $R_{\rm h}$ values of $9.14 \pm 0.01$ arcmin for Collinder 110 and $7.10 \pm 0.01$ arcmin for NGC 2477. Using the angular-to-linear scale listed in Table 2, these angular radii were converted into parsecs (Table 6). The radial distribution of the aggregate mass for both open clusters is illustrated in Figure 10, where $R_{\mathrm{h}}$ is indicated by a red vertical dashed line.

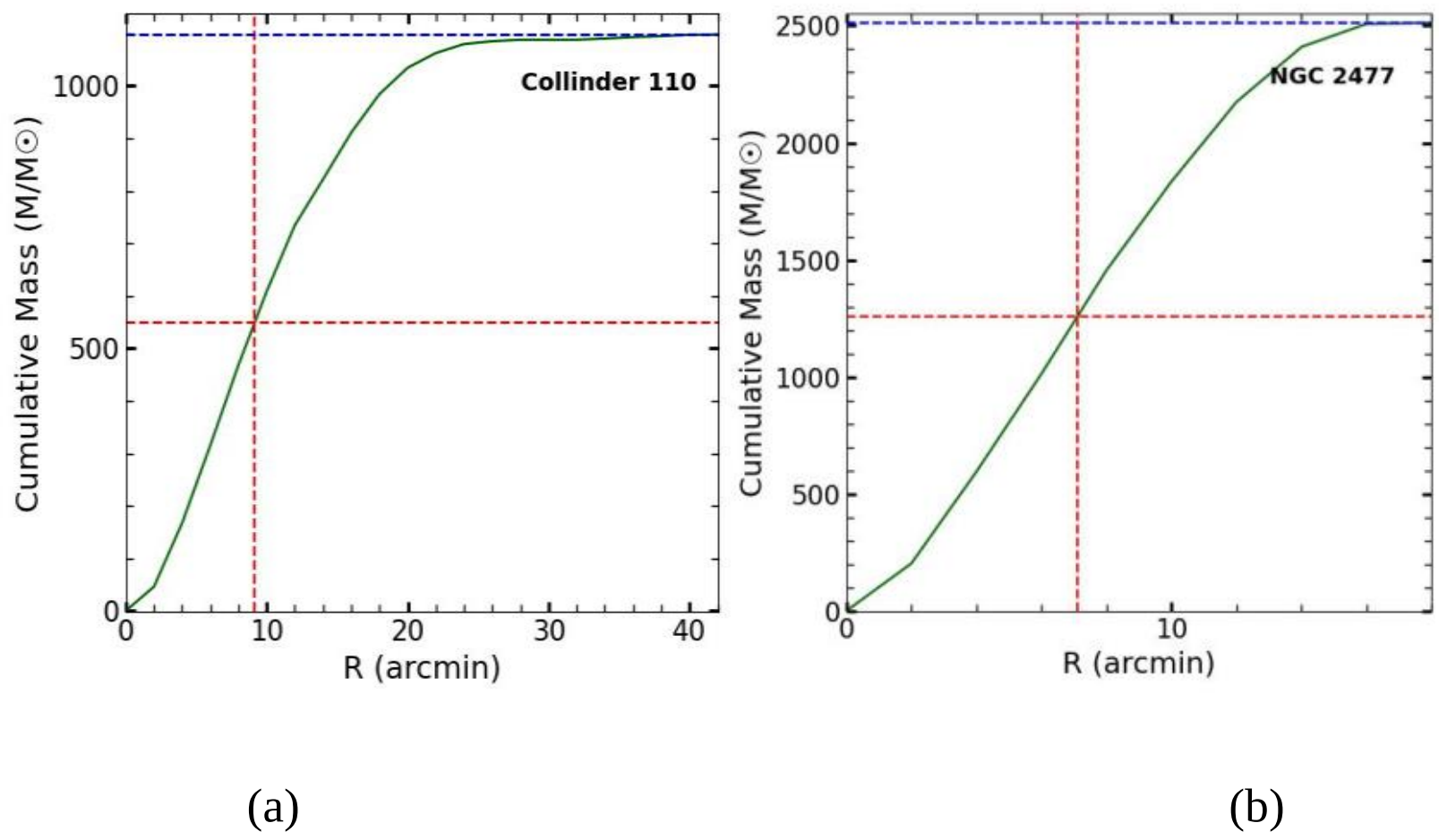

(a) (b)

**Figure 10.** The cumulative mass versus radius of Collinder 110 (panel (a)) and NGC 2477 (panel (b)). The vertical dotted line represents the half-mass radius which corresponds to the half-mass (horizontal red dotted line. The upper blue dotted line denotes the total mass.

Mass segregation, whereby massive stars are preferentially concentrated toward the cluster center, is an important indicator of the dynamical state of an open cluster. We quantified this effect using the mass segregation ratio ($\Lambda_{\mathrm{MSR}}$) following the procedure given by [43]. This analysis included 810 members of Collinder 110 and 2050 members of NGC 2477, using the mathematical formulation provided in Eq. (6). A comprehensive explanation of the variables and physical implications of this equation is available in [43]. The relationship between $\Lambda_{\mathrm{MSR}}$ and $\Lambda_{\mathrm{MST}}$ for both OCs is presented in Figure 11.

$$\Lambda_{\mathrm{MSR}}(N) = \frac{\langle l_{\mathrm{random}} \rangle}{l_{\mathrm{massive}}} \pm \frac{\sigma_{\mathrm{random}}}{l_{\mathrm{massive}}} \quad (6)$$

In Figure 11(a), Collinder 110 displays inverse mass segregation up to $N = 415$. Interestingly, this trend has also been reported in other dynamically evolved systems. Coenda et al. (2025) noted that several old OCs exhibit inverse mass segregation, likely reflecting their long-term dynamical evolution and the influence of external perturbations. NGC 2477 presents mass segregation at all points ($0 < N < 1250$). The significance of the detected mass segregation was assessed using confidence intervals (CI); shown as vertical gray bars in Figure 11, and Anderson-Darling $k$ -sample test ($P_{\mathrm{AD}}$). Up to N=1805, the Anderson–Darling test confirms the presence of mass segregation with $P_{\mathrm{AD}} < 0.05$ throughout this range. Confidence intervals (CI) also confirm the trend except for N=1405, where $\Lambda_{\mathrm{MSR}}$ drops below 1 within the lower confidence interval. For 1805 < N < 2050, both the confidence intervals and the Anderson–Darling test no longer support the presence of mass segregation.

Two additional peaks also show clear signatures of mass segregation, with $\Lambda_{\mathrm{MSR}} = 1.024$ at $N = 650$ and $\Lambda_{\mathrm{MSR}} = 1.027$ at $N = 730$, both supported by their respective confidence intervals. To further validate these findings, we applied the Anderson–Darling $k$-sample test ($P_{\mathrm{AD}}$) [45] as implemented in SciPy [46]. The test yields $P_{\mathrm{AD}}$ values between 0.001 and 0.01 throughout the range $10 < N < 600$, and $P_{\mathrm{AD}} = 0.001$ and $P_{\mathrm{AD}} = 0.019$ for $N = 650$ and $N = 730$, respectively. These results consistently confirm the presence of mass segregation at all the identified points.

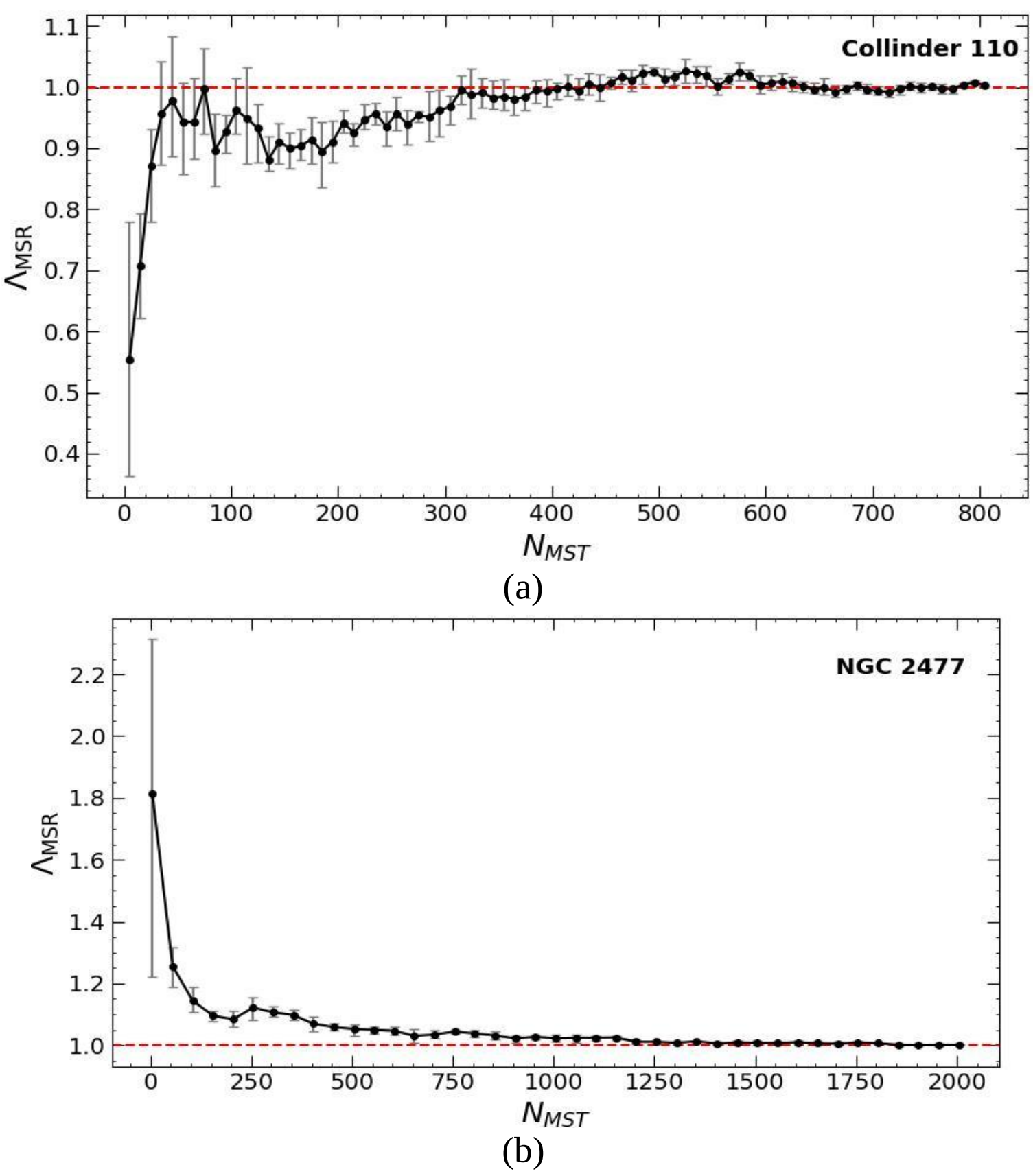

**Figure 11.** Mass segregation ratio ($\Lambda_{MSR}$) as a function of the $N_{\rm MST}$ most massive stars for Collinder 110 (top) and NGC 2477 (bottom). The horizontal red dotted line marks the $\Lambda_{MSR} = 1$ threshold, representing a state of no mass segregation. Gray vertical segments indicate the associated confidence intervals (CI).

In Fig.11(a) Collinder 110 displays inverse mass segregation up to $N = 415$. Interestingly, this trend has also been reported in other dynamically evolved systems. [47] noted that several old OCs exhibit an inverse mass segregation, likely reflecting their long-term dynamical evolution and the influence of external perturbations. NGC 2477 presents mass segregation at all points ($0 < N < 1250$). The significance of mass segregation is tested by confidence intervals (CI); shown in Fig. 11 by the vertical gray bars, and Anderson-Darling $k$ -sample test ($P_{\rm AD}$). Up to $N = 1805$, mass segregation is confirmed by AD test with $P_{\rm AD} < 0.05$ along that the range. Confidence intervals (CI) also confirm the trend except for $N = 1405$, where $\Lambda_{\rm MSR}$ drops below 1 within the lower confidence interval. For $1805 < N < 2050$, both CI and AD test reject the mass segregation.

Two additional peaks also show clear signatures of the mass segregation, with $\Lambda_{\rm MSR} = 1.024$ at $N = 650$ and $\Lambda_{\rm MSR} = 1.027$ at $N = 730$, both supported by their respective confidence intervals. To further validate these findings, we applied the Anderson–Darling $k$-sample test ($P_{\rm AD}$). The test yields $P_{\rm AD}$ values between 0.001 and 0.01 throughout the range $10 < N < 600$, and $P_{\rm AD} = 0.001$ and $P_{\rm AD} = 0.019$ for $N = 650$ and $N = 730$, respectively. These results consistently confirm the presence of mass segregation at all the identified points.

## 6. DYNAMICAL EVOLUTION

The dynamical evolution of open clusters is primarily characterized by the mass function (MF) slope ($\chi$), the degree of mass segregation, the relaxation time ($t_{\rm rlx}$), and the dynamical parameter ($\tau = t/t_{\rm rlx}$) [48]. Two-body relaxation causes massive stars to become more centrally concentrated, while low-mass stars migrate toward the halo and may eventually escape into the Galactic field, leading to progressive mass segregation and depletion of low-mass members. Consequently, clusters with shorter $t_{\rm rlx}$ and larger $\tau$ values are expected to be dynamically more evolved. To evaluate the influence of the Galactic tidal field, we consider the tidal radius ($R_{\rm t}$) and the Jacobi radius ($R_{\rm j}$), where $R_{\rm j}$ marks the distance beyond which the Galactic gravitational force exceeds the cluster's self-gravity [49]. For clusters on nearly circular Galactic orbits, $R_{\rm t}$ is expected to be comparable to $R_{\rm j}$. Following A18, systems with $R_{\rm t} > R_{\rm j}$ are more susceptible to mass loss and disruption, whereas those with $R_{\rm t} < R_{\rm j}$ remain confined within their Roche lobes.

Additional structural indicators include the concentration parameter $c = \log(R_{\rm t}/R_{\rm c})$, which quantifies the degree of central concentration [50,51], the ratio $R_{\rm h}/R_{\rm j}$, which traces the strength of the Galactic tidal influence [4,52,7,8], and $R_{\rm c}/R_{\rm h}$, which measures the compactness of the cluster core.

We define $t_{\rm rlx}$ as the characteristic timescale required for stars to migrate through the cluster via two-body relaxation. Two expressions for $t_{\rm rlx}$ are commonly used in the literature, one based on the limiting radius ($R_{\rm lim}$) and the other based on the half-mass radius ($R_{\rm h}$). The expression based on $R_{\rm lim}$ is given by Eq. (7), and for it usage, see the work of [35].

$$t_{{\rm rlx}_1} = 0.04\left(\frac{N}{\ln N}\right)\left(\frac{R_{\rm lim}}{1\,pc}\right) \tag{7}$$

We adopted $\sigma_{\rm v} \cong 1\ km/s$ as an upper value [54], instead of $\sigma_{\rm v} \cong 3\ km/s$ of [55]. This means that cluster stars move at $\sigma_{\rm v} \cong 1\ km/s$. To quantify the dynamical state of each cluster, we calculated the evolutionary parameter $\tau$ by integrating the estimated cluster ages from Table 4 with their respective relaxation timescales. The mass segregation—characterized by the radial migration of members between the core and the outer regions—is fundamentally governed by $t_{\rm rlx}$ and $\tau$.

The half-mass relaxation or half-light relaxation time $t_{\rm rlx}$ for central parts of the clusters is calculated using the relation given by [56]:

$$t_{{\rm rlx}_2} = \frac{8.9\times10^5\sqrt{N}\times{R_{\rm h}}^{3/2}}{\log(0.4N)\times\sqrt{m}} \tag{8}$$

Here, m refers to the mean mass and N to the number of cluster members. The Jacobi radius ($R_{\rm j}$ ) was calculated using the following equation from [57],

$$R_{\rm j} = \left(\frac{GM_{\rm cl}}{2\varpi^2}\right)^{1/3} \tag{9}$$

where $\varpi$ is the orbital angular velocity, and is calculated by

$$\varpi = \frac{V_\Phi}{R_{\rm GC}} \tag{10}$$

The resulting Jacobi radii are $R_{\rm j} = 16.88 \pm 0.28$ pc for Collinder 110 and $R_{\rm j} = 18.87 \pm 0.07$ pc for NGC 2477, respectively. The concentration parameter $c$ is estimated as $c = 0.77 \pm 0.14$ for Collinder 110 and $c = 0.99 \pm 0.17$ for NGC 2477, respectively. Guiding radii and migration distances were calculated using $R_{\rm guide} = J_{\rm z}/V_\Phi$ and $d_{\rm mig} = R_{\rm guide} - R_{\rm birth}$. The derived dynamical, mass, and migration parameters of Collinder 110 and NGC 2477 are given in Table 6.

Using the metallicity gradient model by [58], we estimated the birth radii ($R_{\rm birth}$) of the two OCs, by utilizing their metal abundances and ages (*Gaia* DR3) given in Table 4.

Initial mass $M_{\rm ini}$ ($M_{\odot}$), ambient density $\rho_{\rm amb}$ ($M_{\odot}pc^{-3}$), and dissolution time $t_{\rm diss}$ ($Myr$) for our sample OCs were calculated using Equations (13)-(15) from [7] and Equation (10) from [8]. Further details are provided in the two references and [59].

The ambient densities ($\rho_{\rm amb}$ ) of our sample OCs were derived from the Galactic potential. $\rho_{\rm amb}$ characterizes the local strength of the Galactic tidal field. The OCs with mass > 0.08 $M_{\odot}$ pc$^{-3}$ are exposed to realtively strong tidal effects, and thus lose their low-mass star content efficiently to the field, whereas the OCs with < 0.08 $M_{\odot}$ pc$^{-3}$ feel these tidal effects less efficently.
The derived values are: $M_{\rm ini} = (7.56 \pm 0.53) \times 10^3 M_{\odot}$, $\rho_{\rm amb} = 0.12 \pm 0.01\ M_{\odot}pc^{-3}$, $t_{\rm diss} = (2.31 \pm 0.08)\ Gyr$ for Collinder 110, and $M_{\rm ini} = (7.83 \pm 0.53) \times 10^3 M_{\odot}$, $\rho_{\rm amb} = 0.12 \pm 0.01\ M_{\odot}pc^{-3}$, $t_{\rm diss} = (2.02 \pm 0.08)\ Gyr$ for NGC 2477, respectively.

**Table 6.** Mass information, dynamical and migration parameters of Collinder 110 and NGC 2477.

| **Parameter** | **Collinder 110** | **NGC 2477** |
|---|---|---|
| Mass range ($M_{\odot}$) | 0.85 – 1.75 | 0.66 – 2.33 |
| MF slope (χ) | 1.23 | 1.84 |
| $R_{\rm h}$ (pc) | 5.47 ± 0.01 | 2.98 ± 0.01 |
| $R_{\rm j}$ (pc) | 16.88 ± 0.05 | 18.87 ± 0.05 |
| $c$ | 0.77 ± 0.14 | 0.99 ± 0.17 |
| $R_{\rm c}/R_{\rm h}$ | 0.85 | 0.79 |
| $R_{\rm h}/R_{\rm j}$ | 0.32 | 0.16 |
| $R_{\rm j}/R_{\rm j}$ | 1.60 | 1.23 |
| $R_{\rm j}/R_{\rm c}$ | 3.65 | 8.03 |
| $R_{\rm guide}$ (kpc) | 10.07 ± 0.02 | 8.69 ± 0.01 |
| $R_{\rm birth}$ (kpc) | 8.84 ± 0.04 | 6.34 ± 0.11 |
| $d_{\rm mig}$ (kpc) | 1.23 ± 0.06 | 2.35 ± 0.12 |
| For $R_{\rm lim}$ | | |
| Cluster mass( $M_{\odot}$) | 1097.30 ± 48.18 | 2512.95 ± 10.42 |
| Mean mass( $M_{\odot}$) | 1.35 ± 0.25 | 1.23 ± 0.44 |
| Members (N) | 810 ± 28 | 2050 ± 45 |
| $t_{\rm rlx_1}$ (Myr) | 119.35 ±10.89 | 238.61 ± 11.95 |
| $\tau_1$ | 15.25 ± 1.62 | 4.39 ± 0.21 |
| $\log \tau_1$ | 1.18 ± 0.05 | 0.64 ± 0.02 |
| $t_{\rm diss}$ (Gyr) | 2.31 ± 0.08 | 2.02 ± 0.08 |
| For $R_{\rm h}$ | | |
| Cluster mass( $M_{\odot}$) | 554.30 ± 3.70 | 1278.71 ± 5.67 |
| Mean mass( $M_{\odot}$) | 0.97 ± 0.01 | 1.31 ± 0.01 |
| Members (N) | 401 ± 20 | 977 ± 31 |
| $t_{\rm rlx_2}$ (Myr) | 104.98 ± 3.68 | 48.24 ± 0.73 |
| $\tau_2$ | 17.34 ± 1.13 | 21.70 ± 0.31 |
| $\log \tau_2$ | 1.24 ± 0.03 | 1.34 ± 0.01 |

## 7. DISCUSSION AND CONCLUSION

Within the uncertainties, the derived reddening values of $E(B-V) = 0.43 \pm 0.05$ mag for Collinder 110 and $E(B-V) = 0.33 \pm 0.04$ mag for NGC 2477 are in good agreement with literature estimates, with differences of $\Delta E(B-V) = 0.03 \pm 0.10$ mag (Table 4). The CMD-based distances ($2.09 \pm 0.06$ kpc and $1.38 \pm 0.04$ kpc) are also consistent, within the uncertainties, with those derived from Gaia DR3 trigonometric parallaxes ($2.08 \pm 0.31$ kpc and $1.45 \pm 0.11$ kpc ; Table 3) [60].

With ages of $1.82 \pm 0.10$ Gyr ($\log Age/yr = 9.26 \pm 0.02$) and $1.05 \pm 0.10$ Gyr ($\log Age/yr) = 9.02 \pm 0.04$), the clusters can be classified as old open clusters, since old open clusters are generally defined as having ages greater than 700 Myr ($\log Age/yr > 8.85$). Our derived ages are also consistent, within the uncertainties, with literature values (Table 4). The small-to-moderate discrepancies in age and distance modulus (or distance) may arise from the use of solar-metallicity isochrones for clusters spanning the metallicity range [Fe/H] = [−0.13, +0.11] (Col. 4 of Table 4), differences in reddening determinations from CMDs or CC diagrams, and the adoption of different membership-selection techniques, as also emphasized by Karataş et al. (2023).

Collinder 110 and NGC 2477 are located in the third Galactic quadrant, outside the solar circle, with ($\ell$, $R_{\rm GC}$)= (210°, 10.07 kpc) and ($\ell$, $R_{\rm GC}$) = (254°, 8.69 kpc), respectively. Both clusters lie close to the Galactic plane, with maximum vertical distances of $z_{\rm max} = 0.23$ and $0.15$ pc (Table 5). They have completed approximately 6 and 9 revolutions around the Galactic center, respectively (Table 5). During their lifetimes, they have completed approximately 6 and 9 revolutions around the Galactic center (Table 5), implying prolonged exposure to external tidal perturbations that may contribute to their gradual dissolution.

Their relatively steep mass function slopes ($\chi = 1.23$ and $1.84$) indicate that low-mass stars dominate over massive ones. Because their ages are significantly greater than their relaxation times ($t_{\rm rlx_1} = 119$ Myr and $t_{\rm rlx_1} = 239$ Myr, Table 6), both clusters can be considered dynamically relaxed. The evolutionary parameter of NGC 2477 ($\tau_1 = 4.39$) suggests a modest level of dynamical evolution. In contrast, the larger evolutionary parameter of Collinder 110 ($\tau_1 = 15.25$), together with its steep mass function, indicates advanced dynamical evolution and possible preferential loss of low-mass stars from its outer regions due to external tidal effects. In this context, Collinder 110 exhibits signs of mild mass segregation. Figure 11 further supports this interpretation: both clusters show evidence of mass segregation, reaching peak $\Lambda_{\rm MSR}$ values of 1.30, 1.35, and 1.15 in the ranges $N < 5$, $10 < N < 20$, and $25 < N < 40$, respectively.

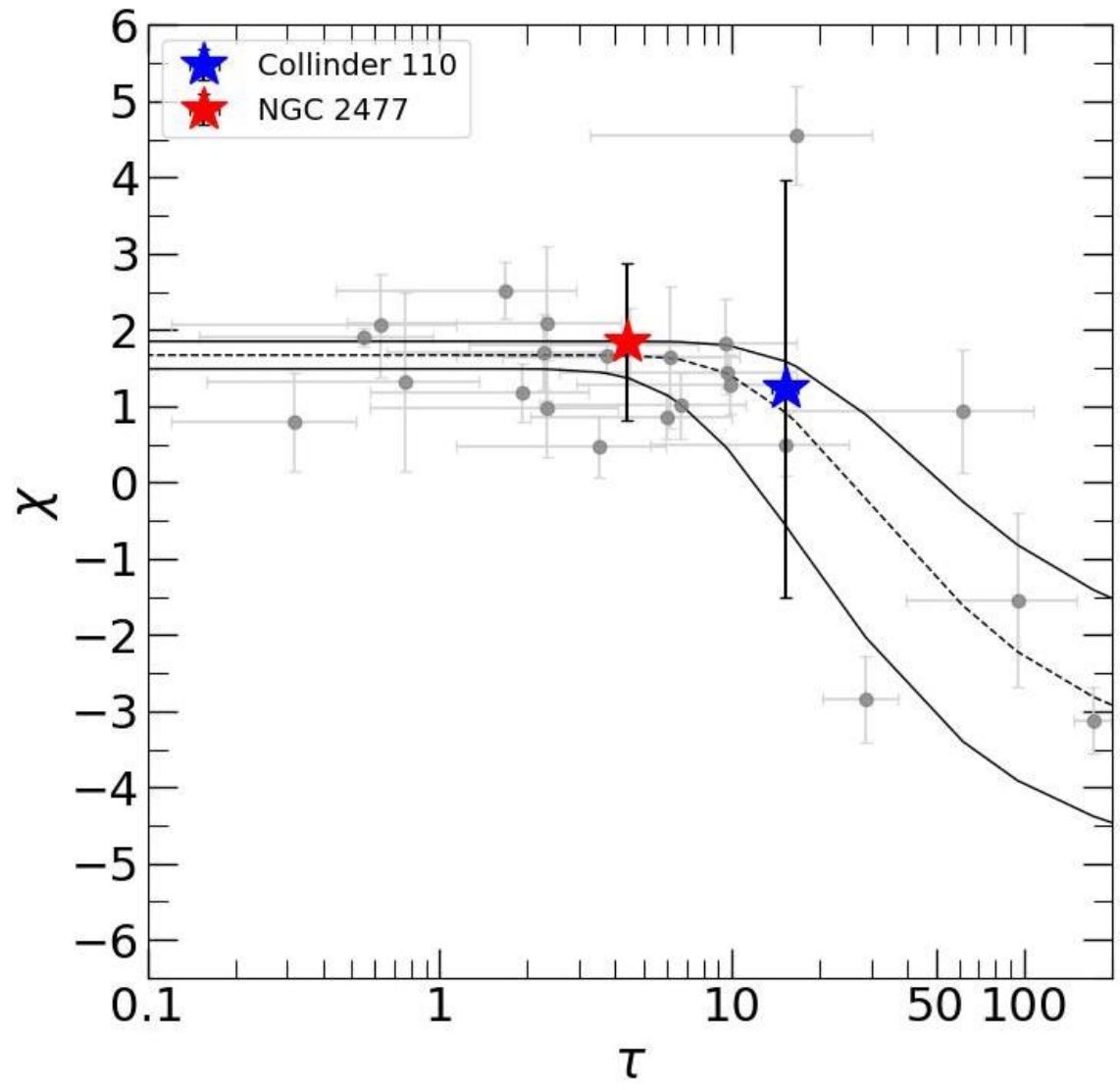

**Figure 12.** The loci of Collinder 110 and NGC 2477 in $\chi$ versus $\tau$. The grey dots with the 1σ uncertainty and solid curves show the 24 OCs of G17. The dotted line shows the fit, $\chi(\tau) = \chi_o - \chi_1 e^{-\tau_o/\tau}$. The overall MF slope undergoes an exponential decay with $\tau$.

Figure 12 shows that the global mass function (MF) slope of Collinder 110, associated with a high dynamical evolution parameter ($\tau_1 = 15.25$), approaches zero and even becomes negative. This trend likely reflects the preferential depletion of low-mass stars through internal two-body interactions and the cumulative impact of external Galactic tidal forces. Conversely, NGC 2477 presents comparatively flat MF slopes with $\tau_1 = 4.29$, suggesting that it is at an earlier stage of dynamical evolution. As reported in [53], noticeable MF flattening typically occurs for systems with $\tau \lesssim 7$; however, this threshold depends on the characteristics of the analyzed cluster sample, particularly the balance between young and old objects. Two different formulations of relaxation time are adopted in this work. The analysis presented in Figure 12 is based on the global relaxation time $t_{\rm rlx,1}$(Equation 7), derived using the limiting radius $R_{\rm lim}$. In contrast, Figure 14 employs the half-mass relaxation time $t_{\rm rlx,2}$ (see Section 6.1).

The evolutionary status of Collinder 110 and NGC 2477 is further examined using the dynamical criteria described in [61, 62, 63, 7, 8], as shown in Figures 13 and 14. Following G17, benchmark values of $2000\, M_{\odot}$ and $(R_{\rm c}, R_{\rm lim}) = (1.5, 7)$ pc are adopted to categorize clusters in terms of mass and structural scale.

Both clusters exhibit relatively extended structures: $(R_{\rm c}, R_{\rm lim}) = (4.63, 24.67)$ pc for Collinder 110 and $(2.35, 22.19)$ pc for NGC 2477. In the $R_{\rm c}$–$R_{\rm lim}$ diagram (Figure 13a), they generally follow the expected positive correlation, although minor offsets are evident. Their large limiting radii are indicative of substantial stellar evaporation. These structural properties may result from gradual mass segregation and may also reflect their initial formation conditions. Additionally, dynamical heating by stellar-mass black holes could have contributed to the outward redistribution of stars, potentially enhancing mass loss toward the field [64]. Both clusters are situated in the third Galactic quadrant, a region relatively poor in giant molecular clouds, which may partly account for their survival over 1.82 Gyr and 1.05 Gyr, respectively. Despite continuous dynamical evolution, their overall MF slopes remain positive, implying that their original low-mass stellar populations were sufficiently abundant.

In the age–radius relations (Figure 13b–c), Collinder 110 lies well beyond the bifurcation age of ~1 Gyr, while NGC 2477 is positioned near this transition. This suggests that core expansion has likely influenced the enlargement of their limiting radii. The dependence of cluster size on age provides insight into open-cluster survival and disruption timescales [62], with the bifurcation age near 1 Gyr identified in Galactic samples by [62, 63]. Taking their dissolution times into account ($t_{\rm diss} = 2.31$ Gyr for Collinder 110 and 1.98 Gyr for NGC 2477; Table 6), Collinder 110 has completed roughly 79% of its expected lifetime and lost approximately 15% of its initial mass ($7560\, M_{\odot}$). NGC 2477, on the other hand, has lost about 33% of its original mass ($7630\, M_{\odot}$) and reached nearly 53% of its predicted dissolution timescale. Thus, although both clusters are dynamically evolved and structurally extended, they have endured significant tidal influences, likely aided by their locations in the Galactic disk.

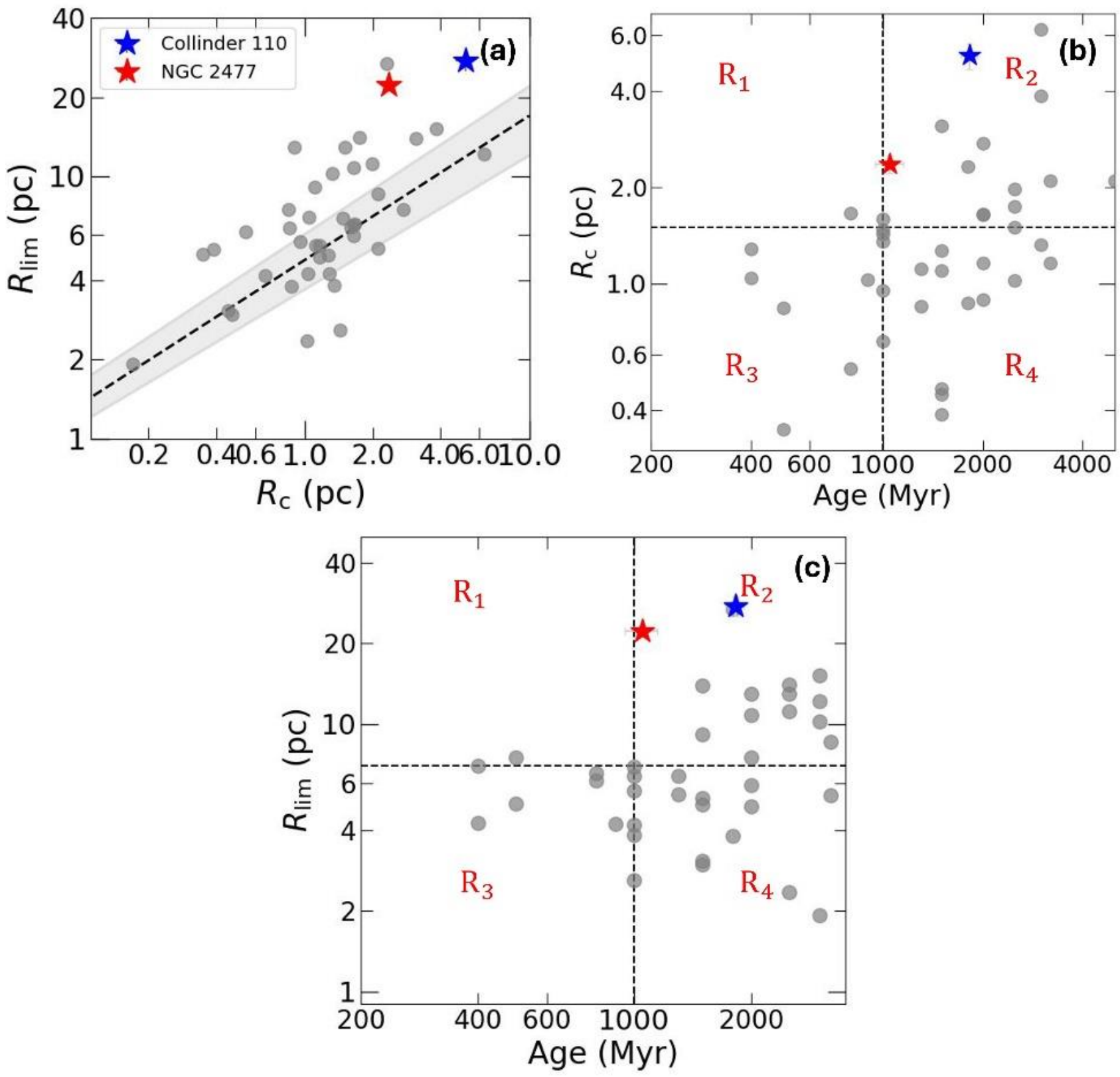

**Figure 13.** Positions of Collinder 110 (blue star) and NGC 2477 (red star) in the structural parameter diagrams. (a) The ($R_{lim} - R_c$) relation, where the dashed line shows the best-fit relation and the grey shaded region represents the 1σ uncertainty. (b–c) Variations of ($R_c$) and ($R_{lim}$) with cluster age, respectively. The regions $R_1$–$R_4$ are indicated in panels (b) and (c), while the horizontal and vertical dotted lines separate relatively small and large open clusters. Grey dots represent the comparison sample of open clusters from [61].

The $R_c$ and log $\tau_2$ values of both OCs, together with the measurements reported by [8] (gray dots), are displayed in Figure 14a. [7] and [8] reported a decreasing trend in core radius after $\tau \sim 2.2$ for the OCs with $R_{GC} < 7$ kpc (see their figures 8-11). However, for our clusters, which are located at larger Galactocentric distances, it is difficult to conclude that the core radius of Collinder 110 has begun to decrease although it has advanced evolutionary time (log $\tau_2$=1.24). In contrast, the core radius of NGC 2477 remains consistent with the reported trend despite its advanced dynamical age ($\log \tau_2 = 1.34$). While our $R_c$ values (Table 2) are consistent with the values reported by [7] (5.43 pc, 2.63 pc), their evolutionary times ($\log \tau = 0.17$ and 0.27) are smaller than ours ($t_{rlx_2}$ values in Table 6). Their large half-mass relaxation times ($t_{rlx} = 1778$ Myr and 605 Myr) using Equaton (8) result from their large numbers of cluster members ($N$), large $R_h$ and high cluster mass (see [7]' tables 1-2). The same interpretation applies to panels e and f in Fig. 14. Note that the ages of this study and [7] are in concordance within the error limits. In this respect, discrepancies among different authors regarding the number of cluster members, total cluster mass and cluster sizes can lead to different interpretations of the dynamical evolution.

Clusters characterized by $R_t/R_j > 1$ are considered tidally overextended, while systems with $R_t/R_j < 1$ remain gravitationally confined within their Jacobi radius and retain most of their stellar content [7]. Based on this criterion, both OCs with $R_t/R_j = 1.60$ and 1.23 are exposed to external tidal effects, and have therefore preferentially lost low-mass stars to the Galactic field. Their ($R_h$, $R_c$) values also satisfy the condition $R_h > R_c$ (Figure 14c). The ratios of ($R_c/R_h$, $R_h/R_j$) = (0.85, 0.32) (Collinder 110) and ($R_c/R_h$, $R_h/R_j$) $= (0.79, 0.016)$ (NGC 2477) agree better with the locations of the sample OCs of [8] (panel~d).

Following the arguments of [8] and [4], the core radii of some OCs shrink during advanced dynamical evolution, as a consequence of mass segregation and evaporation of less massive stars in their core regions. According to [8], the core radius $R_c$ is less affected by external environmental influences compared to $R_h$ and $R_j$, indicating that the structural properties of the central region are primarily governed by internal dynamical processes rather than Galactic tidal effects.

Collinder 110 and NGC 2477 with $c = 0.77$ and $0.99$ are highly concentrated and possess dynamically relaxed internal structures, and they obey negative correlation between $c$ and $R_h/R_j$ (Figure 14b). Their central parts are in advanced dynamical evolution with large evolutionary times $\log \tau_2 = 1.24$ and $1.34$. Their $R_h/R_j$ tends to decrease with increasing $\log \tau_2$ as a consequence of the internal relaxation process (Figure 14e), which causes the clusters' main body to be progressively more compact. With their relatively small $R_h/R_j$ ratios of $0.32$ and $0.16$, these OCs have allowed their internal mass distributions to relax across their Jacobi radii $R_j = 16.88$ pc and $18.87$ pc (the point of being tidally filling) without being tidally disrupted ($R_h/R_j < 0.70$). They are affected by Galactic tidal forces, and thus have preferentially lost low-mass stars to the field since their ratios fall in the range $0.12 < R_h/R_j < 0.35$, according to figure 2a of [4].

In panel (d) the $R_c/R_h$ ratios of [7] are somewhat smaller than our measurements. Note that the determination of $R_h$ of this study and theirs is different. In panel~(d) the dynamical evolution of both OCs is ruled by both internal and external dynamic processes. The large $R_c/R_h$ ratios of both OCs are indicative of two-body relaxation, mass segregation and the possible presence of stellar-mass black holes in their central regions.

With the $\rho_{\rm amb} = 0.12$ $M_\odot$ pc$^{-3}$ ($\log \rho_{\rm amb} = -0.92$), our sample OCs which are located outside the solar circle ($R_{GC} > 8$ kpc) occupy the region $R_h/R_j < 0.40$ in [8]'s figure 10b. Both OCs have more extended Roche lobes (See Table 6 for $R_j$ values), therefore exhibit relatively small $R_h/R_j$ ratios in terms of their large $R_h$ values. The two OCs experience the effects of the external tidal fields to some extent. [8] shows that open clusters located in environments with $\log \rho_{\rm amb} < -1.1$ experience lower evaporation rates, since their dynamical evolution is mainly driven by internal relaxation processes. In regions with low ambient density, the Jacobi radius $R_j$ increases as the surrounding density decreases, leading to a reduction in the $R_h/R_j$ ratio. Consequently, tidal influences from the Galactic environment become comparatively weaker.

Our results indicate that Collinder 110 and NGC 2477 have migrated approximately 1.23 kpc and 2.35 kpc from their birth locations, respectively. This displacement suggests that their present guiding radii differ significantly from their original positions (Table 6). Such substantial radial migration is consistent with enhanced mass loss and advanced dynamical evolution, as reflected by their large $\log \tau_2$ values and the depletion of low-mass stars.

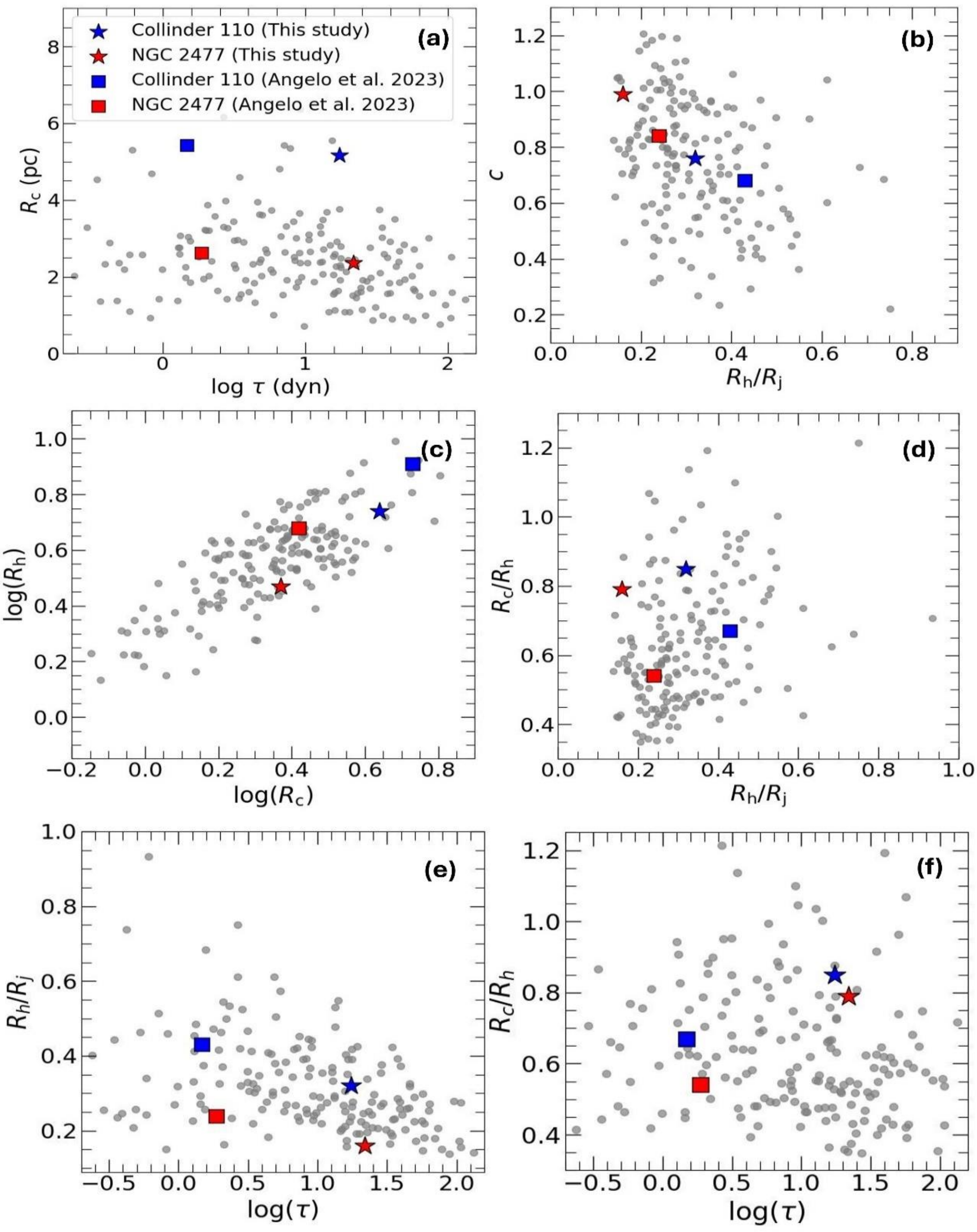

**Figure 14.** Positions of Collinder 110 (blue star) and NGC 2477 (red star) in the structural and dynamical parameter relations. (a) $R_c$ versus $log\,\tau$; (b) concentration parameter (c) versus $R_h/R_j$; (c) $log\,R_h$ versus $log\,R_c$; (d) $R_c/R_h$ versus $R_h/R_j$; (e) $R_h/R_j$ versus $log\,\tau$; and (f) $R_c/R_h$ versus $log\,\tau$. Grey dots represent the comparison sample from [8], while blue and red squares indicate the corresponding values reported by [7].

## ACKNOWLEDGMENTS

Zahra AL gratefully acknowledges the financial support provided by İlim Yayma Vakfı (İYV), Türkiye. The author also thanks the anonymous reviewers for their careful reading and constructive comments, which helped improve the quality and clarity of this manuscript.

## CONFLICT OF INTEREST

The authors stated that there are no conflicts of interest regarding the publication of this article.

## CRediT AUTHOR STATEMENT

**Zahra Al:** Conceptualization, Methodology, Software, Formal analysis, Investigation, Writing – original draft, Visualization. **Yüksel Karataş:** Conceptualization, Methodology, Formal analysis, Validation, Writing – review & editing, Supervision. **Forough Rajaei:** Software, Formal analysis, Investigation, Data Curation.